\documentclass[journal,comsoc]{IEEEtran}
\usepackage{cite}
\usepackage{amsmath,amssymb,amsfonts}
\usepackage{graphicx}
\usepackage{textcomp}

\usepackage{framed,multirow}
\usepackage{url}
\usepackage{xcolor}
\usepackage{lineno,hyperref}
\definecolor{newcolor}{rgb}{.8,.349,.1}

\usepackage{bm}
\usepackage{dsfont}
\usepackage{algorithm}
\usepackage{algorithmicx}
\usepackage{algcompatible}
\usepackage{pifont}
\usepackage{threeparttable}
\usepackage{makecell}
\usepackage{rotating}

\usepackage[normalem]{ulem}

\usepackage{marvosym}
\graphicspath{{img/}}
\DeclareGraphicsExtensions{.pdf,.png}
\usepackage[caption=false,font=small]{subfig}

\begin{document}
\title{AG-CRC: Anatomy-Guided Colorectal Cancer Segmentation with Imperfect Anatomical Knowledge}
\author{Rongzhao Zhang, Zhian Bai, Ruoying Yu, Wenrao Pang, Lingyun Wang, Lifeng Zhu, Xiaofan Zhang, Huan Zhang, and Weiguo Hu
	\thanks{Equal contribution: R. Z. and Z. B.}
	\thanks{R. Z., W. P. and X. Z. are with Shanghai Artificial Intelligence Laboratory, Shanghai, China. Z. B., R. Y., L. Z. and W. H. are with the Shanghai Digital Medicine Innovation Center, Ruijin Hospital, Shanghai Jiao Tong University School of Medicine, Shanghai, China. L. W. and H. Z. are with the Department of Radiology, Ruijin Hospital, Shanghai Jiao Tong University School of Medicine, Shanghai, China. W. H. is also with the Department of Surgery, Ruijin Hospital, Shanghai Jiao Tong University School of Medicine, Shanghai, China.}
	\thanks{Corresponding authors: X. Z. (e-mail: zhangxiaofan@pjlab.org.cn), H. Z. (email: huanzhangy@163.com ) and W. H. (email: wghu@rjh.com.cn).}}

\maketitle

\begin{abstract}
When delineating lesions from medical images, a human expert can always keep in mind the anatomical knowledge behind the voxels. However, although high-quality (though not perfect) anatomical information can be retrieved from CT scans with modern deep learning algorithms, it is still an open problem how these automatically generated organ masks can assist in addressing challenging lesion segmentation tasks, such as the segmentation of colorectal cancer (CRC). In this paper, we develop a novel Anatomy-Guided segmentation framework to exploit the auto-generated organ masks to aid CRC segmentation from CT, namely AG-CRC. First, we obtain multi-organ segmentation (MOS) masks with existing MOS models (e.g., TotalSegmentor) and further derive an organ of interest mask that robustly covers most of the colon-rectum and CRC voxels. Then, we propose an anatomy-guided training patch sampling strategy by optimizing a heuristic gain function that considers both the proximity to regions of interest and sample diversity. Third, we design a novel self-supervised learning scheme inspired by the topology of tubular organs like the colon to boost the model performance further. Finally, we define an anatomy-inspired loss function to focalize the model training process. We extensively evaluate the proposed method on two CRC segmentation datasets, where substantial performance improvement (5\%$\sim$9\% in Dice) is achieved over current state-of-the-art medical image segmentation models, and the ablation studies further evidence the efficacy of every proposed component.
\end{abstract}


\section{Introduction}
\label{sec:introduction}
\IEEEPARstart{C}{olorectal} cancer (CRC) is a significant global health concern, which ranks as the third most commonly diagnosed cancer and the second leading cause of cancer death worldwide in 2020 \cite{sung2021global}. 
Non-invasive medical imaging, particularly computed tomography (CT), plays a pivotal role in the early detection, accurate diagnosis, and effective treatment of CRC \cite{argiles2020localised}. However, the manual interpretation of CT scans for colorectal tumors is time-consuming, subjective, and prone to inter-observer variability. To address these challenges, the automated colorectal cancer segmentation technique is a promising direction. It provides essential information on tumor size, location, and shape, which may facilitate the diagnosis, treatment planning, and monitoring of disease progression.

While CRC segmentation from CT scans holds immense potential for improving diagnosis and treatment, it is very challenging to achieve accurate and reliable segmentation results, partially due to the large variability of the tumor morphology and the proximity of the colorectal region to neighboring organs, such as the stomach, small intestine, and bladder, etc. To this date, the most prevailing and powerful approach to addressing medical image analysis tasks, including CRC segmentation, might be the deep learning (DL) technique \cite{ronneberger2015u,isensee2021nnu,huang2018hl,panic2020convolutional}, which feature the ability to learn complex patterns and representations automatically from large-scale medical data with proper annotation. Nevertheless, most existing DL-based medical image segmentation algorithms tackle the problem in a pure image-to-label manner without considering prior knowledge of anatomical structures, such as the location, type, and shape of neighboring organs.

When reading medical imaging scans like CT, it is natural and necessary for a radiologist to keep in mind which organ s/he is looking at. However, unless manual delineation of multiple organs is performed in advance, existing DL-based models rarely have the ability to utilize the knowledge of anatomical structures. As a consequence, the lack of anatomical knowledge may lead to false positive detection outside the target organ, and more importantly, the training process may also be hindered as the model have to spare its attention to the nuisance outside the organ of interest. Some previous works \cite{ye2021anatomy,yao2022deepcrc} attempt to augment the segmentation model by providing extra manual organ annotations, but this leads to additional annotation labor and drastically increases the development cost.

\newcommand{\picwidthMOS}{0.8in}
\begin{figure}[]
	\captionsetup[subfigure]{labelformat=empty}
	\centering
	\subfloat{\includegraphics[width=\picwidthMOS]{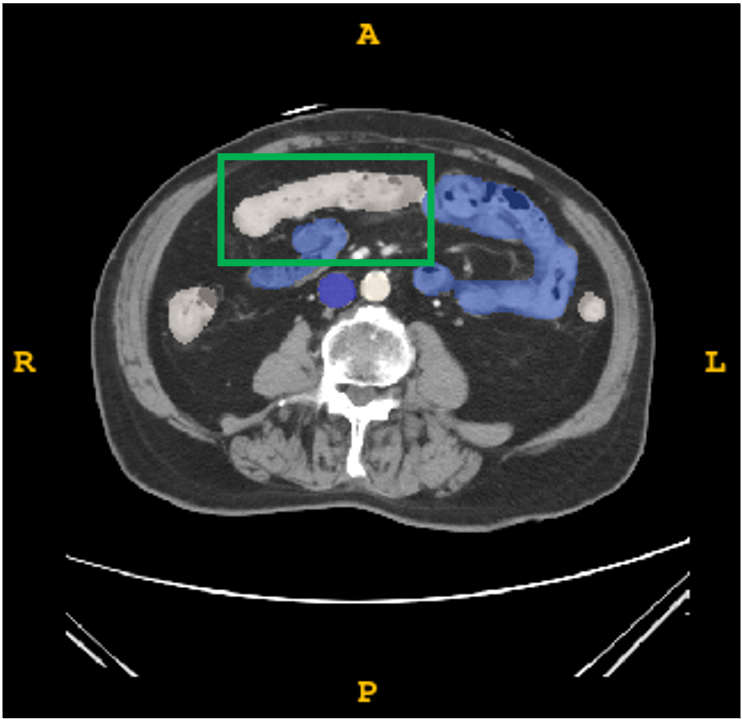}}
	\hspace{0in}
	\subfloat{\includegraphics[width=\picwidthMOS]{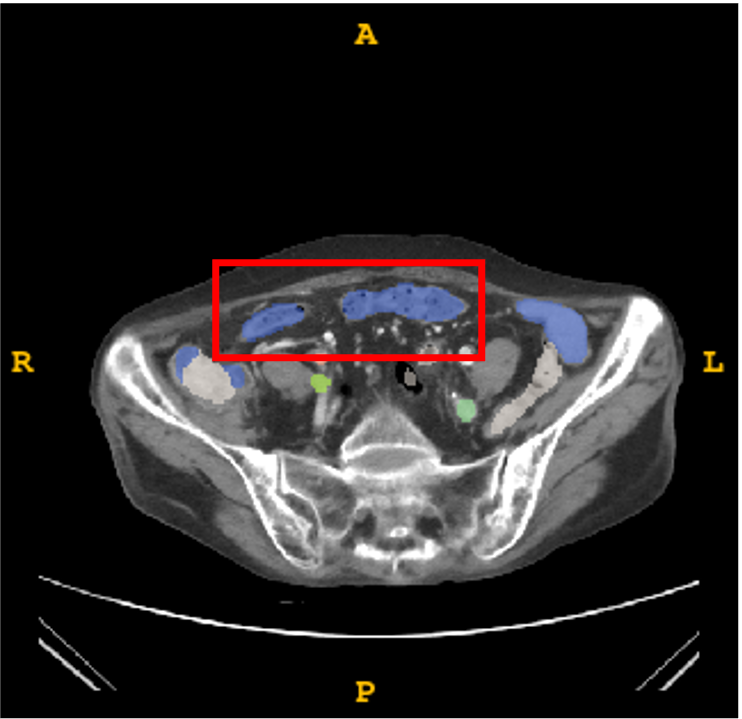}}
	\hspace{0in}
	\subfloat{\includegraphics[width=\picwidthMOS]{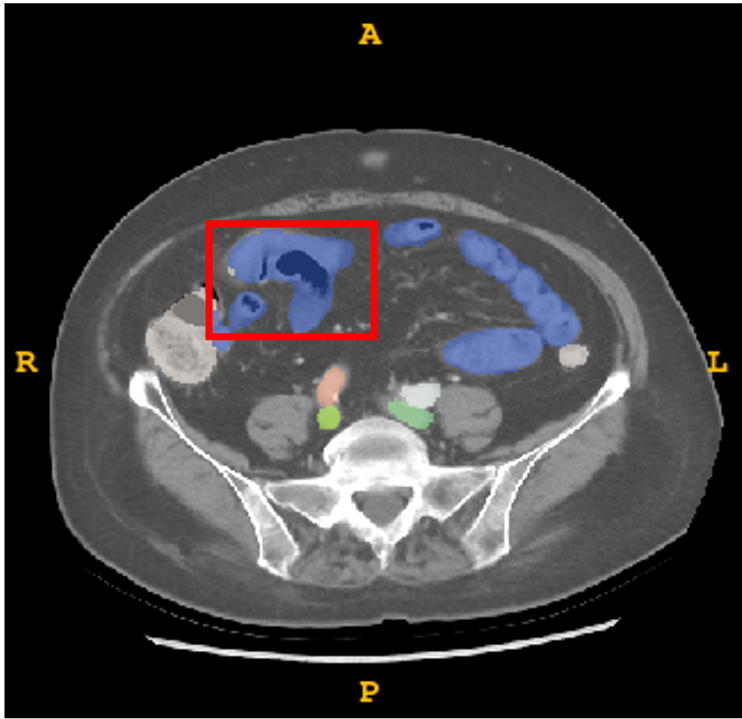}}
	\hspace{0in}	
	\subfloat{\includegraphics[width=\picwidthMOS]{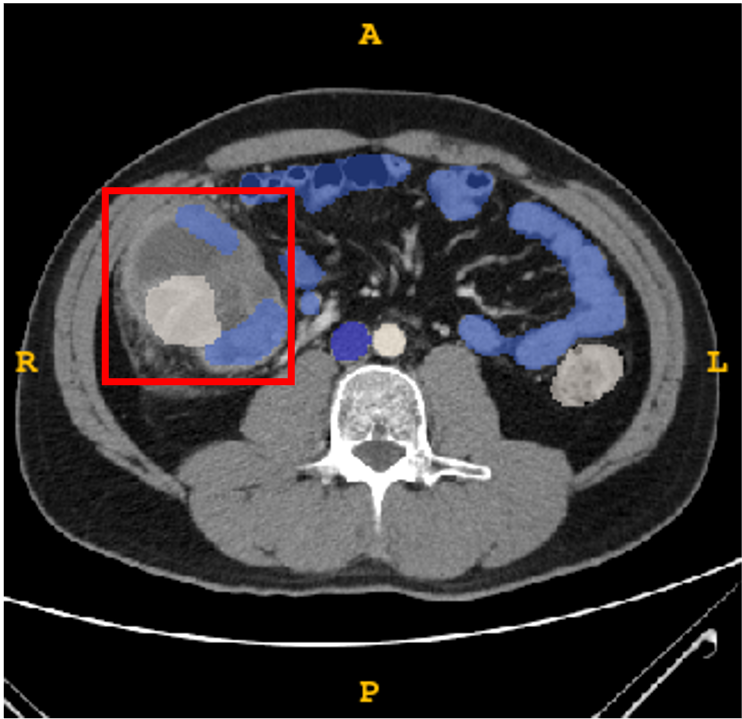}}
	
	\subfloat[(a) Correct]{\includegraphics[width=\picwidthMOS]{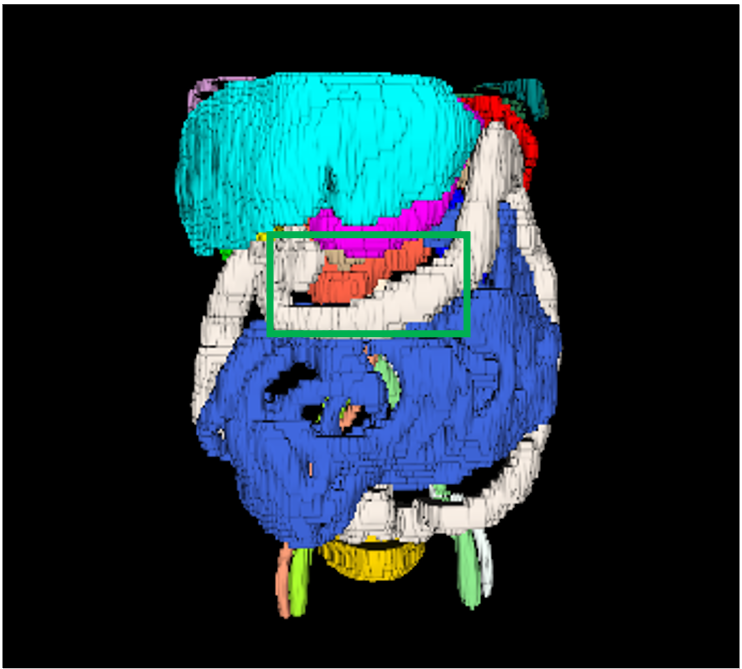}}
	\hspace{0in}	
	\subfloat[(b) Discontinue]{\includegraphics[width=\picwidthMOS]{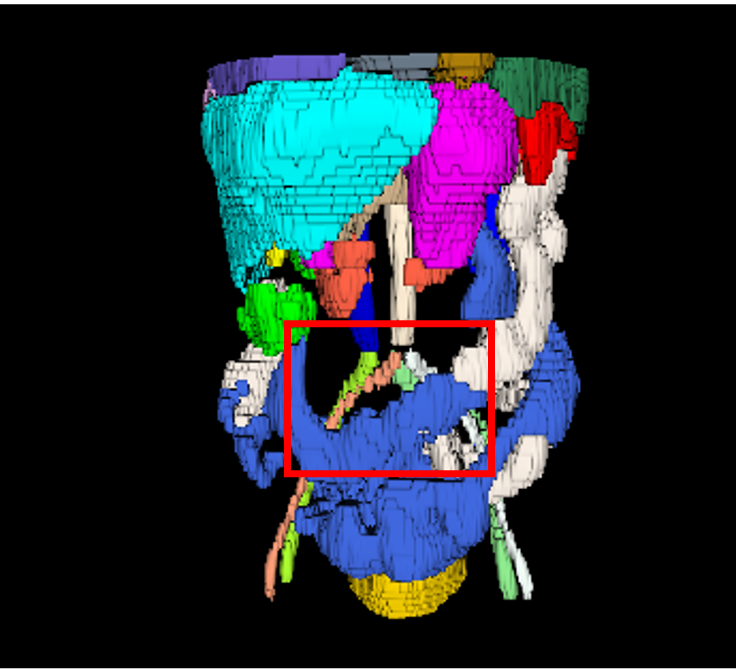}}
	\hspace{0in}	
	\subfloat[(c) Discontinue]{\includegraphics[width=\picwidthMOS]{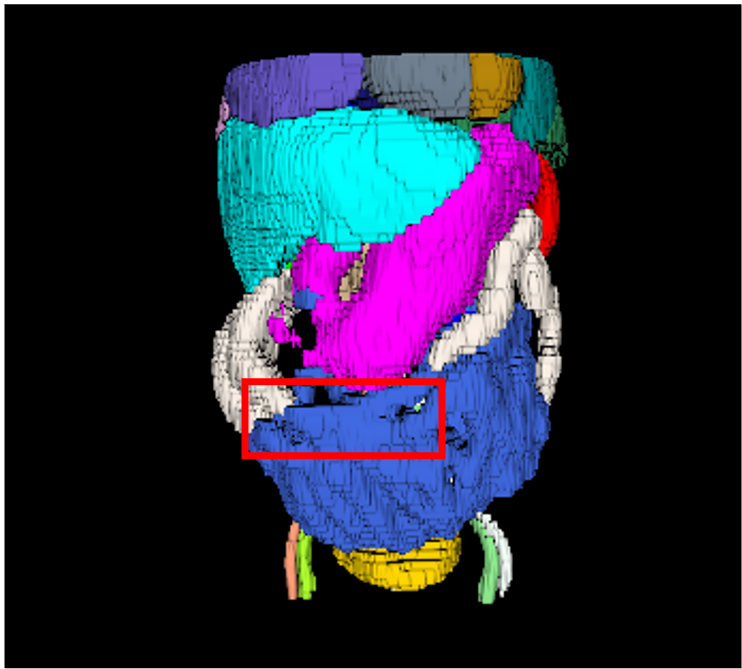}}
	\hspace{0in}	
	\subfloat[(d) Missing]{\includegraphics[width=\picwidthMOS]{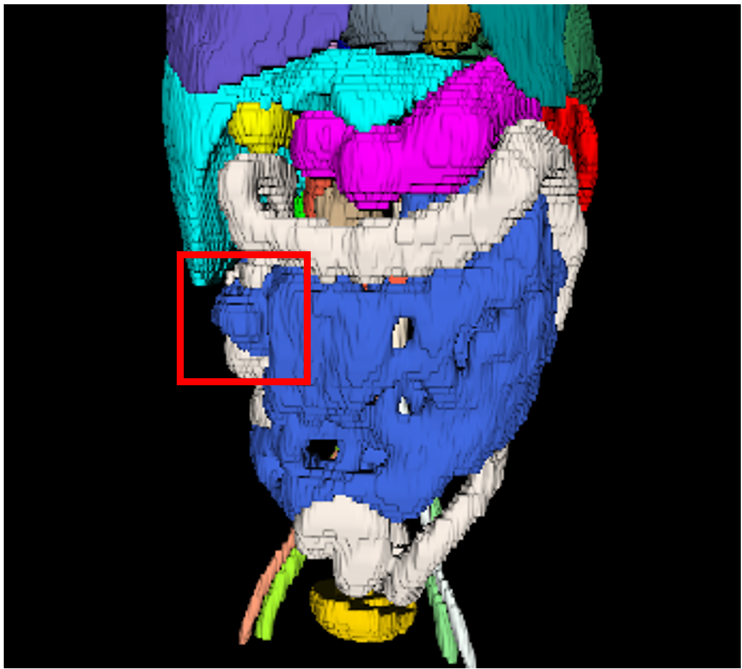}}
	
	\caption{Examples of imperfect auto-generated MOS masks. The first row illustrates axial CT slices overlapped by organ masks, and the second row shows the corresponding 3D reconstruction of organs. Different organs are distinguished by colors. Especially, the colon-rectum is gray and the intestine is in blue. Column (a) demonstrates a correct case without obvious organ segmentation errors (a green box highlights the correct continuity of the transverse colon), (b) and (c) show two cases with confusion between colon (gray) and small intestine (blue), where the discontinuity of colon is highlighted by red bounding boxes, (d) shows a case where the segmentation has missed considerable voxels due to pathology (a large colon tumor), as highlighted by red boxes.}
	\label{fig_mos}
\end{figure}
Recently, with the fast evolution of medical image segmentation algorithms and the release of abdominal multi-organ segmentation (MOS) datasets \cite{landman2015miccai,wasserthal2022totalsegmentator,luo2022word}, MOS algorithms gradually become mature and are now reaching a point where the predicted organ masks are usable for downstream tasks, such as lesion segmentation in certain organs. Nonetheless, although state-of-the-art MOS algorithms can delineate major abdominal organs with a satisfying accuracy in theory, e.g., \cite{luo2022word} reports an average Dice of 0.87 for colon segmentation on the WORD dataset with nnUNet \cite{isensee2021nnu}, noticeable failure still exists (such as discontinuity of the intestinal tract, confusion among similar organs like colon and small intestine, etc., as illustrated in Fig. \ref{fig_mos}), especially when the CT scans are from different medical centers or the anatomical structures are heavily distorted due to pathology. As there are few preceding works that have successfully integrated anatomical knowledge into DL models without additional annotation effort, and existing MOS models' prediction is still flawed, it is an open problem how the imperfect auto-generated anatomical information can be leveraged to boost segmentation performance on challenging lesion tasks like CRC segmentation. 

In this paper, we propose an anatomy-guided CRC segmentation framework that can effectively utilize the imperfect anatomical knowledge generated by automatic MOS algorithms, called AG-CRC. In this framework, the anatomical information is seamlessly embedded in each of the four stages in a neural network training pipeline: data preparation, training patch sampling, self-supervised pertaining, and supervised learning. Concretely, in the data preparation stage, we apply a state-of-the-art MOS model, i.e., TotalSegmentor \cite{wasserthal2022totalsegmentator}, on the CT scans to obtain the multi-organ segmentation (MOS), and then we derive an expanded gastro-intestinal region based on the MOS mask, called the organs of interest (OOI) mask, which compensates for the segmentation error (especially the missing voxels) of the MOS mask in colorectum. Second, to determine what the model ``sees'' during the learning phase, we compute a probabilistic sampling map (PSM) by optimizing a heuristic gain function that takes into account both the voxel relevancy and data diversity. Third, to encourage the model to learn more relevant features during the pretraining phase, we design a novel anatomy-inspired proxy task, where the bowel wall region is masked for reconstruction rather than random rectangular blocks. Lastly, during the supervised learning stage, we developed a masked loss function based on the OOI map so that the model could avoid sparing its attention to the disturbing features outside the target organs, thus, the segmentation performance could be further boosted. We extensively evaluated the proposed framework on two CRC datasets, i.e., the public MSD-Colon dataset (126 cases) and an in-house dataset containing 388 patients, and compared to a wide range of state-of-the-art 3D medical image segmentation algorithms. The results indicate that our framework outperforms these competitors by a large margin on the CRC segmentation task, and detailed ablation studies on the collected dataset also corroborate the efficacy of each anatomy-enhanced component. 

In summary, the main contributions of this work are four-fold: 
\begin{itemize}
	\item We propose a novel anatomy-guided CRC segmentation framework that can efficiently utilize both the CT scan and the automatically derived anatomical information. To the best of our knowledge, this is the first work that exploits auto-generated multi-organ masks to enhance the performance of a colorectal cancer segmentation network.
	\item 
	Several novel anatomy-guided schemes are developed to enhance the model training process, including anatomy-augmented input modalities, a novel sampling strategy guided by anatomical structures, a more clinically relevant self-supervised learning scheme and an anatomy-inspired loss function. 
	
	\item The proposed CRC segmentation framework is extensively evaluated on two CRC datasets, where it demonstrate superior performance over state-of-the-art models by a large margin (e.g., 5\%$\sim$9\% in Dice on the two datasets), and significant improvement brought by each proposed component can be observed in the detailed ablation study.
	
	\item To facilitate and encourage more anatomy-guided research in the medical image computing community, we will release a derived dataset that is composed of the generated multi-organ masks and the expanded gastro-intestinal region masks (i.e., the MOS and OOI masks) on the four public datasets (i.e., MSD-Colon, LiTS, WORD and TCIA-Lymph) used in this work\footnote{Both the derived dataset and the code will be available at \url{https://github.com/rongzhao-zhang/AG-CRC}}. 
\end{itemize}

\section{Related works}
\subsection{Medical image segmentation}
With the prosperity of neural network models, deep-learning-based approaches have been the predominant direction in the medical image segmentation (MIS) community \cite{shen2017deep}. 
Among the many deep models developed for various MIS tasks, UNet \cite{ronneberger2015u} and its variants \cite{cciccek20163d,isensee2021nnu} stood out due to its excellence in performance and elegant design of architecture. As an example, nnUNet~\cite{isensee2021nnu}, a popular framework based on the UNet concept, presents a rule-based self-configuring framework that emits state-of-the-art performance on a vast range of MIS tasks. More recently, with the surge of transformer models \cite{vaswani2017attention} in natural vision fields \cite{dosovitskiy2020image,liu2021swin}, a number of transformer-based MIS algorithms are also invented, such CoTr \cite{xie2021cotr}, SwinUNETR \cite{tang2022self}, nnFormer \cite{zhou2023nnformer}, etc. Nevertheless, although some transformer-based approaches can achieve better results than nnUNet on certain datasets, nnUNet is still one of the most popular frameworks for MIS due to its high performance and great generalizability.

\subsection{Colorectal cancer segmentation}
Many efforts have been devoted to the challenging colorectal cancer segmentation problem from medical imaging data such as CT and MRI. For example, \cite{wang2018deep} applies 2D UNet to perform tumor segmentation on rectal MR images, \cite{panic2020convolutional} employs a semi-automatic patch-based algorithm that combines T2w, DWI and ADC modalities to augment the performance and \cite{jiang2021ala} proposes a sophisticated lesion-aware 3D CNN to deal with CRC segmentation from pelvic T2w MRI. Compared with MRI, CT imaging is more economic and has larger scanning range than pelvic MR, which is deemed as a better imaging modality for tubular organs like colon and rectum \cite{yao2022deepcrc}. Hence, the segmentation of CRC from CT scans may have better applicability than MR in clinical trials. To identify CRC from CT, \cite{liu2019accurate} combines a 2D FCN with a generative adversarial network (GAN) to refine the probabilistic prediction maps, but their dataset only contains selected 2D slices rather than the whole CT volume. \cite{yao2022deepcrc} builds a novel colorectal coordinate system to enhance the CRC segmentation model, but this system requires extra annotation of the colon and rectum, and manual intervention is required to extract the centerline of the colorectum.

\subsection{Anatomy-guided medical image processing}
Integrating anatomical information into medical image processing algorithms has been an attractive topic since the early stage of medical image analysis (MIA) research. For example, statistical shape models (SSMs) \cite{lorenz2000generation,heimann2009statistical} capture the inherent shape prior and variability of anatomical structures within a population; atlas-based methods \cite{cuadra2004atlas,iglesias2015multi} propagate pre-computed organ/tissue atlases to target images to align labels with pixels. These methods ensure anatomical correctness by resorting to pre-defined atlas or shape priors, which also limits their expressive power.
In the era of deep learning, many researches attempt to combine the strength of deep models and conventional anatomy-aware MIA algorithms. 
\cite{hoogi2016adaptive} employs CNN to provide the initial lesion location and to compute the parameters of the active contour functional; DALS \cite{hatamizadeh2019deep} introduces a multi-scale CNN to learn the initialization probability map for the active contour model; 
\cite{li2021agmb} fuses the features of anatomical landmarks and X-ray to emit better evaluation results; \cite{zhou2021anatomy} combines domain adaption, organ segmentation, and surface-point-based registration to obtain a better CT to MRI registration algorithm; \cite{jiang2022anatomy} leverages anatomy knowledge by separating gyri and sulci (two folding types of cerebral cortex) nodes and building a bipartite graph network. 
Nevertheless, although these methods have achieved performance improvement over their baselines, most of them exploit anatomical knowledge in an ad-hoc manner and are usually task-specific, making it less suitable to other problems like CRC segmentation from 3D CT. 

\begin{figure*}[]
	\centering
	\includegraphics[width=0.95\textwidth]{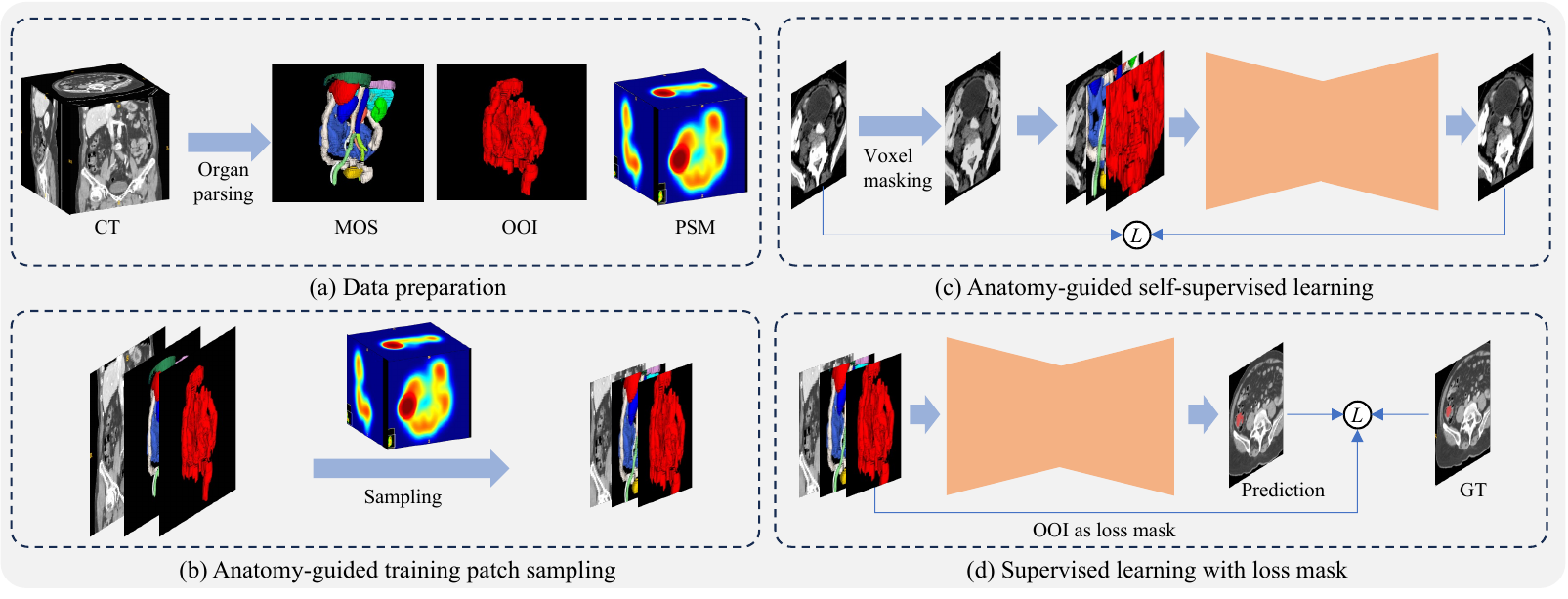}%
	\caption{Overview of the proposed framework. The (imperfect) anatomical knowledge is first obtained during data preparation and then injected into the training framework in three phases: patch sampling, self-supervised pertaining, and supervised learning. }
	\label{fig_overview}
\end{figure*}
\section{Methodology}
\subsection{Overview}
In Fig. \ref{fig_overview}, we schematically depict the proposed anatomy-guided CRC segmentation framework. First, during data preparation, the multi-organ segmentation (MOS) masks are generated, and the organs of interest (OOI) masks are derived; second, a probabilistic sampling map (PSM) is computed based on the OOI and the tumor locations for training patch sampling; third, a novel voxel masking scheme is developed to make the pretraining phase anatomy-aware; lastly, an anatomy-focalizing loss is applied during the supervised learning phase to further boost model performance. Each of the proposed components will be elaborated in the following sections. 

\subsection{Generation of organ masks}\label{sec_mask_gen}
To retrieve the anatomical information of an abdominal CT scan, we employ TotalSegmentor \cite{wasserthal2022totalsegmentator}, a state-of-the-art MOS model, to segment major abdominal organs such as spleen, kidneys, liver, stomach, pancreas, duodenum, intestine, colon, etc. However, although the predicted organ masks have a good overall quality, they still have erroneous outputs in some key structures, such as the confusion between the small intestine and colon, and voxels may be missing due to large appearance variation or pathology. Some examples can be found in Fig. \ref{fig_mos}. To overcome the imperfect organ segmentation and hint the model about the organs of interest, we derive an expanded binary organ mask by merging the organs in the gastrointestinal tract, including the stomach, duodenum, small intestine, colon, and rectum, as these organs resemble colon in both location and appearance. Besides, to cover those missing voxels like case (d) in Fig. \ref{fig_mos}, we employ morphological operations to further dilate the organ mask. Since the expanded organ mask contains the organs of interest for CRC segmentation, we called it the OOI (organs of interest) mask. To improve robustness, we generate a second OOI mask with another MOS algorithm, i.e., the WORD \cite{luo2022word} model, and then merge the two OOIs by binary OR operations. The overall algorithm for generating the organ masks (MOS and OOI) is detailed in Algorithm \ref{algo_ooi}. 

To utilize the derived MOS and OOI masks, we concatenate them with the CT scan and feed them (as a 3-channel volume) into the segmentation model. Moreover, the OOI mask plays a central role in our segmentation framework, based on which several anatomy-aware schemes are proposed, as introduced in the following sections.

\begin{algorithm}[]
	\caption{Organ mask generation}
	\label{algo_ooi}
	\begin{algorithmic}[1]
		\STATEx \textbf{Input:} 
		\STATE \quad CT scan $X$, 
		\STATE \quad TotalSegmentor model $M_{TS}$, 
		\STATE \quad WORD model $M_{WORD}$,
		\STATE \quad Organ indicator sets for TotalSegmentor ($S_{TS}$) and WORD ($S_{WORD}$), indicating the integer labels corresponding to the organs of interest in the predicted organ masks, respectively,
		\STATE \quad Dilation repeat times $t$.
		\STATEx \textbf{Output:} one MOS mask (from $M_{TS}$), one OOI mask
		
		\STATEx {\textbf{Generation of multi-organ masks}}
		\STATE \quad Evaluate $Y_{TS} \leftarrow M_{TS}(X)$
		\STATE \quad Evaluate $Y_{WORD} \leftarrow M_{WORD}(X)$
		
		\STATEx {\textbf{Generation of OOI mask}}
		\STATE \quad  $O_{TS} \leftarrow \text{zeros\_like}(Y_{TS})$
		\STATE \quad {for $v$ in $S_{TS}$:}
		\STATE \quad \quad $O_{TS} \leftarrow (Y_{TS} == v) + O_{TS}$
		\STATE \quad $O_{TS} \leftarrow O_{TS} > 0$
		\STATE \quad Repeat for $t$ times:
		\STATE \quad \quad $O_{TS} \leftarrow \text{binary\_dilation}(O_{TS})$
		\STATE \quad Obtain $O_{WORD}$ by repeating line 8 to line 13 using $Y_{WORD}, O_{WORD}, S_{WORD}$
		\STATE \quad $O_{final} \leftarrow \text{binary\_or}(O_{WORD}, O_{TS})$ 
		\STATE \textbf{Return} $Y_{TS}, O_{final}$
		
	\end{algorithmic}
\end{algorithm}

\subsection{Anatomy-aware training patch sampling}
Given the fact that CRC only exists in the colorectal region, which is a subset of the OOI, the voxels that are far away from the OOI region might be simple negatives that contribute little to the training process. Based on this observation, we believe that an optimal sampling strategy should consider both the proximity to high-value regions (e.g., the OOI region) and the diversity of sampled patches. Hence, we propose to obtain the optimal probabilistic sampling map (PSM) by maximizing a gain function that takes these factors into account, called anatomy-guided sampling (AG-sampling). Concretely, the gain function at location $i$ is:
\begin{equation}\label{eq_g}
	g_i = K_i\cdot O,
\end{equation}
where $K_i$ is the patch kernel at location $i$, $O$ is a binary map indicating the voxels of interest, e.g., the OOI mask, $A\cdot B$ denotes the inner product of $A$ and $B$. The dimension of $K_i$ and $O$ are the same as the input CT scan $X$. Given the $i$-th location $\bm p_i=(x_i,y_i,z_i)^T$, the patch kernel $K_i$ can be defined as a truncated Gaussian function:
\begin{equation}\label{eq_k}
	K_i(\bm q) = \left\{
	\begin{array}{ccl}
		C\exp\left(-\frac{1}{2}\left(\bm q - \bm p_i\right)^T\bm\Sigma^{-1}\left(\bm q - \bm p_i\right)\right),&{\bm q \in Q_i}\\
		0, &{\bm q \notin Q_i}
	\end{array} \right.
\end{equation}
where $C = (2\pi)^{-3/2}\det(\bm\Sigma)^{-1/2}$ is the normalizing constant, $Q_i = \{\bm x : \|\bm x - \bm p_i\| \preccurlyeq \bm d / 2\}$ ($\preccurlyeq$ denotes element-wise less than or equal to) is the $\bm d$-size neighborhood of $\bm p_i$, $\bm d=(d_x, d_y, d_z)^T$ denotes the training patch size, and $\bm\Sigma = \text{diag}(\bm \sigma)$ is the covariance matrix with $\bm \sigma = 0.1\bm d$.

To maximize the overall gain function, we optimize the following function
\begin{equation}\label{eq_sampling}
	\begin{array}{cl}
		\underset{S}{\min} & - \sum_i{s_ig_i} + \frac{1}{2}\mu\sigma^2(S) \\
		\text { s.t. } \quad & s_i \ge 0 \quad \forall i, \\
		& \sum_i s_i = 1,
	\end{array}
\end{equation}
where $S=\{s_i\}_{i=1}^{n}$ is the probabilistic sampling map (PSM) to be optimized, $\sigma^2(\cdot)$ denotes the variance function, $\mu$ is a balancing coefficient (we simply use $\mu=1$). In the objective, the first term encourages a high probability for high-gain location, while the second term accounts for the diversity of the sampling by penalizing a large variance of PSM's distribution. And the constraints ensure a valid probability distribution. By looking into the problem (\ref{eq_sampling}), we notice that the solution to its unconstrained version is close-form:
\begin{equation}\label{eq_solution}
	\hat s_i = \frac{1}{\mu}g_i + \bar s,
\end{equation}
where $\bar s$ denotes the average value of $S$, here we let $\bar s=1 / n$ and $n$ is the number of elements in $S$. As the solution (\ref{eq_solution}) is both easy-to-acquire and meaningful, instead of solving (\ref{eq_sampling}) exactly, we choose to solve it approximately by directly projecting the intermediate solution (\ref{eq_solution}) to the feasible region by
\begin{equation}
	s_i = \frac{\hat s_i}{\sum_i\hat s_i},
\end{equation}
such that the constraints can be satisfied. In this way, the gain optimization problem (\ref{eq_sampling}) can be solved with high efficiency. 

As for the selection of high-value regions, we notice that the positions of the key organs and the tumor are both valuable clues for patch sampling. To combine both, we first obtain two PSMs $S_{organ}$ and $S_{tumor}$ by setting $O$ as the OOI mask and the ground truth tumor mask, respectively, and then merge them to get the final sampling map:
\begin{equation}\label{eq_smpl_combine}
	S_{final} = (1-\lambda)S_{organ} + \lambda S_{tumor},
\end{equation}
where we set $\lambda=0.33$, striking a balance between the tumor and the relevant organs.

\subsection{Anatomy-guided self-supervision}\label{sec_ssl}
As the colon and rectum are both tubular organs with a hollow structure, colorectal tumors usually grow on the bowel walls. Based on this characteristic of CRC, we assume that it would be beneficial for a model to get familiar with the appearance of bowel walls in its pretraining stage. Therefore, we propose a novel anatomy-guided self-supervision learning (AG-SSL) task for the segmentation model, which masks out the \textit{bowel wall} region and requests the model reconstruct it. 

\begin{figure*}[]
	\centering
	\includegraphics[width=0.95\textwidth]{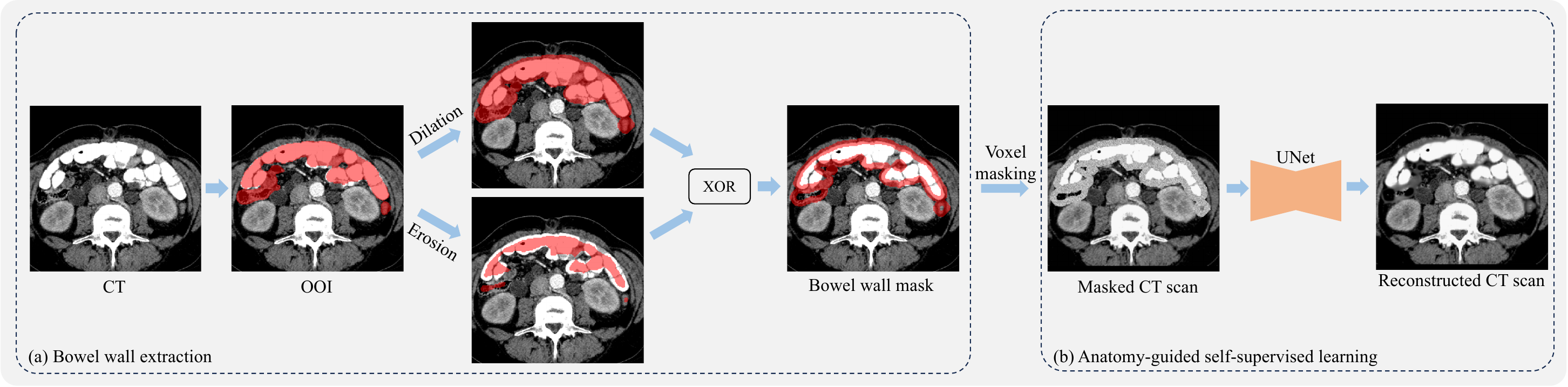}%
	\caption{Bowel wall mask generation and the anatomy-guided SSL task. The bowel wall masks are generated offline prior to the pretraining. During SSL, the UNet also takes as input the two extra data modalities, i.e., MOS and OOI masks, which are omitted here for simplicity.}
	\label{fig_ssl}
\end{figure*}

To retrieve the bowel wall region in a CT scan, we design a morphology-based extraction method, as illustrated in Fig. \ref{fig_ssl}(a). First, we generate the OOI masks of a CT scan, following the method described in \S\ref{sec_mask_gen}\footnote{Here, we ignore the dilation step in OOI generation, as we need a more precise border to extract bowel walls.}. Then, we perform binary dilation and erosion operations on the OOI mask, respectively, and use the XOR operation to compute the difference between the dilated and the eroded regions, resulting in the ``thickened edges'' of the OOI mask (last column of Fig. \ref{fig_ssl}(a)).  As these edges can cover most boundaries between bowel content and abdominal environment, we employ them as the bowel wall mask, denoted as $B$. 

Given the extracted bowel walls, we define the AG-SSL task as:  
\begin{equation}\label{eq_ssl}
	\begin{array}{cl}
		\underset{W}{\min} & \left|\left|X - f\left(\hat X, O, M; W\right)\right|\right|_1\\
		\text { s.t. } \quad & \hat X = \mathcal{M}(X, B),
	\end{array}
\end{equation}
where $X$ is the original CT scan, $\mathcal{M}(X, B)$ denotes a function that substitutes the voxels of image $X$ within the bowel wall region $B$ by Gaussian noises, $O$ and $M$ are the corresponding OOI and MOS masks, respectively, which are concatenated with the masked CT image as model inputs. Overall, AG-SSL first masks out the bowel walls by random values and then train the model to reconstruct the CT scan in an auto-encoder fashion with $L_1$ loss function, as illustrated in Fig. \ref{fig_ssl}(b).  With this proxy task, we hope that the model could be aware of the general appearance of the bowel wall, (normally) a thin tissue separating bowel content and abdominal environment, which is the key anatomical structure in the CRC segmentation task. 

\subsection{Anatomy-focalized loss function}
As colorectal cancer only happens in colorectal region, it is a natural idea to focus model training only on the colorectal region. Nevertheless, given the imperfect nature of the auto-generated organ masks, as described in \S\ref{sec_mask_gen} and Fig. \ref{fig_mos}, simply filtering model outputs by the imperfect colorectal mask may incur severe false negatives. To address this challenge, we propose to apply the OOI masks derived in \S\ref{sec_mask_gen} to focalize the training process. 
Formally, we define an anatomy-focalized loss (AF-loss) function as
\begin{equation}\label{eq_loss}
	L_M(Y, \hat Y, O) = L_0(Y,\hat Y \circ O),
\end{equation}
where $L_0$ denotes the conventional segmentation loss, which is the combination of a Dice loss and a cross-entropy loss, $Y$ denotes the ground truth (GT) label, and $\hat Y$ is the predicted soft tumor mask, $O$ is the binary OOI mask, $\circ$ denotes element-wise multiplication. Intuitively, this loss function ignores all the non-OOI area during training, which makes the supervised learning phase only attending to the OOI region that may contain the target tumors. 

\section{Experimentation}
\subsection{Colorectal cancer datasets}
To evaluate the proposed method, we collect a large colorectal cancer CT dataset from Shanghai Ruijin Hospital (Shanghai, China), which consists of 388 patients diagnosed with colorectal cancer. For each patient, a contrast-enhanced abdominal CT scan (venous phase) is collected, and the tumor region is annotated by an experienced gastroenterologist and later verified by another senior radiologist. During the annotation phase, the physicians are also provided with the corresponding post-surgery pathological report to narrow down the search area for the tumor(s). The age of the involved patients ranges from 15 to 91, and the CT scans were acquired from July 2013 to July 2022, 52.32\% of them are male, and 47.68\% are female. According to AJCC Cancer Staging Manual \cite{amin2017eighth}, the pathologic cancer staging (T for tumor and N for lymph node) statistics of the collected dataset are summarized in Table \ref{tab_t} and \ref{tab_n}, respectively. 
The tumor size ranges from 0.84 to 617.17 cm$^3$, with a median size of 32.33 cm$^3$. Most of the cases (374 out of 388) only involve a single tumor, and the other 14 cases have two CRC lesions each. All the scans share the same in-plane dimension of $512\times 512$, and the dimension along the z-axis ranges from 36 to 146, with a median of 91. The in-plane spacing ranges from $0.60\times 0.60$ to $0.98\times 0.98$ mm, with a median of $0.76\times 0.76$ mm, and the z-axis spacing is from 5.0 to 7.5 mm, with a median of 5.0 mm. We split the whole dataset into three splits, with 238, 30, and 120 for training, validation, and testing, respectively. We refer to this dataset as the RJH-Colon dataset in the following sections. This study has been approved by the Institutional Reviewer Board of the Ruijin Hospital, Shanghai Jiao Tong University School of Medicine.

\begin{table}[!htb]
	\begin{minipage}{0.5\linewidth}
		\caption{Statistics of T-staging}
		\label{tab_t}
		\centering
		\begin{tabular}{llllll}
			\hline
			T-staging & T1                     & T2                     & T3                     & T4                      \\
			Count     & \multicolumn{1}{r}{31} & \multicolumn{1}{r}{46} & \multicolumn{1}{r}{83} & \multicolumn{1}{r}{228} \\ \hline
		\end{tabular}
	\end{minipage}%
	\begin{minipage}{0.5\linewidth}
		\centering
		\caption{Statistics of N-staging}
		\label{tab_n}
		\begin{tabular}{llll}
			\hline
			N-staging & N0                      & N1                     & N2                     \\
			Count     & \multicolumn{1}{r}{277} & \multicolumn{1}{r}{75} & \multicolumn{1}{r}{36} \\ \hline
		\end{tabular}
	\end{minipage} 
\end{table}

Besides the collected datasets, we also validate the proposed algorithm on a public dataset, i.e., the colon cancer dataset in the Medical Segmentation Decathlon (MSD) challenge \cite{antonelli2022medical}. This dataset releases 126 pairs of CT scans and colon tumor annotation. The CT scans are acquired at Memorial Sloan Kettering Cancer Center (New York, NY, USA) and are of portal venous phase. The slice thickness of this dataset ranges from 1.25 to 7.5 mm, with a median spacing of $5.0\times 0.78\times 0.78$ mm. Since this dataset is relatively small, we conduct standard 5-fold cross-validation on this dataset. We refer to this dataset as the MSD-Colon dataset in the following sections.

\subsection{Pretraining datasets}
To perform self-supervised learning, we collected three publicly available abdominal CT datasets, including a CT Lymph Nodes dataset \cite{Roth_Summers_2015} from The Cancer Imaging Archive (TCIA) \cite{clark2013cancer}, a liver tumor dataset (LiTS) \cite{bilic2023liver} and a multi-organ segmentation dataset (WORD) \cite{luo2022word}. In total, these datasets contain 527 contrast-enhanced abdominal CT scans from multiple centers across countries, including USA (TCIA-Lymph, 176 cases), China (WORD, 150 cases), and international project (LiTS, 201 cases). The slice thickness of these data ranges from 0.45 to 6.0 mm, with a median spacing of $1.25\times 0.86\times 0.86$ mm. Among the 527 CT scans, we employ 500 of them for self-supervised learning and keep the rest 27 scans for quality examination of the pre-trained model.

\subsection{Network architecture and system implementation}
The network architecture is automatically determined by the nnUNet framework given our dataset and a specified target spacing $5.0\times 0.78\times 0.78$ mm, resulting in a deep UNet with 32 convolution or transposed convolution layers (including the auxiliary deep supervision heads) and consists of 31M parameters in total. 

We implement the algorithm with Python 3.10 and PyTorch 2.0. The experiments are conducted on multiple sites. One example hardware configuration is a PC with a 32-core i9 13900K CPU, 64GB RAM, and one NVIDIA RTX 4090 GPU with 24GB VRAM. A typical training time (for 1000 epochs) is 12 hours, and the VRAM consumption is less than 9GB. 

\subsection{Evaluation metrics}
To thoroughly evaluate the involved models, we employ the Dice similarity coefficient (Dice), normalized surface Dice (NSD), precision (PR), recall (RE) and 95-th percentile Hausdorff distance (HD95) to measure the quality of model predictions. For NSD, the tolerant value is set to 4 mm following the practice of the MSD challenge \cite{antonelli2022medical}. For HD95, as there usually exist some volumes with an empty prediction, where the HD95 metric may go to infinity or NaN values, we set their HD95 scores as 1000 mm to penalize the failure while making the averaged score still meaningful. 

\section{Results}
To demonstrate the superiority of the proposed anatomy-guided CRC segmentation algorithm, we first conduct head-to-head comparisons with several state-of-the-art 3D medical image segmentation models. Then, we carry out extensive ablation studies to analyze the effectiveness of each proposed component. Finally, we visualize some representative qualitative results. 

\subsection{Comparison with state-of-the-art methods}
We compare the proposed framework with five state-of-the-art 3D medical image segmentation models, including nnFormer \cite{zhou2023nnformer}, UNETR++ \cite{shaker2022unetr++} CoTr \cite{xie2021cotr}, SwinUNETR \cite{tang2022self} and nnUNet \cite{isensee2021nnu}, where nnFormer and UNETR++ are purely transformer-based, whose encoder and decoder are both built by transformer blocks; CoTr inserts a deformable transformer block between the CNN-based encoder and decoder; SwinUNETR is a hybrid model that employs a shifted-window transformer as encoder and a CNN as decoder; nnUNet is purely a CNN model. All these methods are carefully adapted to the CRC segmentation task following their released source code. In addition, as SwinUNETR also supplies a pretrained model by self-supervised learning, we report its results both with and without pretraining. 

The comparison results with existing state-of-the-art segmentation models are listed in Table \ref{tab_comp_msd} and Table \ref{tab_comp_srh} for the MSD-Colon and RJH datasets, respectively. Our framework achieves the highest Dice, NSD, recall and the lowest HD95 distance on both datasets, which outperforms the second-best (nnUNet or SwinUNETR) by a large margin. For example, on the MSD-Colon dataset, our method achieves an overall Dice of 0.5086, outperforming the second-best nnUNet by more than 5\%. Similarly, on the RJH dataset, an average Dice of 0.6935 is observed for the proposed method, which is a 9\% improvement compared with nnUNet. This trend holds on other measurements. Considerable performance improvement ranging from 4\% to 8\% can be observed on NSD for both datasets, which further verifies the strong and positive effect brought by our framework. Since our method is developed based on the nnUNet framework, the nnUNet can be regarded as the baseline model that does not interact with the auto-generated anatomical knowledge, hence the large performance improvement should be attributed to the incorporation of the additional anatomical information and the novel and effective way to utilize them. The in-depth ablation studies and analysis in the following sections will further reveal how much each proposed components contributes to the model performance.

\begin{table}[]
	\caption{Comparison with state-of-the-art medical image segmentation algorithms on the MSD-Colon dataset.}
	\label{tab_comp_msd}
	\centering
	\begin{tabular}{lrrrrr}
		\hline
		Model            & \multicolumn{1}{c}{Dice$\uparrow$}  & \multicolumn{1}{c}{PR$\uparrow$} & \multicolumn{1}{c}{RE$\uparrow$} & \multicolumn{1}{c}{NSD$\uparrow$} & \multicolumn{1}{c}{HD95$\downarrow$}   \\ \hline
		nnFormer         & 0.3742                   & 0.3824                        & 0.4700                     & 0.4131                  & 195.10                   \\
		UNETR++          & 0.3142                   & 0.5483                        & 0.3021                     & 0.3624                  & 326.98                   \\
		CoTr             & 0.3735                   & 0.6119                        & 0.3652                     & 0.4305                  & 270.86                   \\
		SwinUNETR        & 0.4383                   & 0.5051                        & 0.4688                     & 0.4979                  & 140.20                   \\
		SwinUNETR w/ SSL & 0.4457                   & 0.5327                        & 0.4728                     & 0.5075                  & 135.27                   \\
		nnUNet           & 0.4559                   & \textbf{0.6330}               & 0.4739                     & 0.5204                  & 238.04                   \\
		Ours             & \textbf{0.5086}          & 0.5907                        & \textbf{0.5294}            & \textbf{0.5804}         & \textbf{125.15}          \\ \hline
	\end{tabular}
\end{table}

\begin{table}[]
	\caption{Comparison with state-of-the-art medical image segmentation algorithms on the RJH-Colon dataset.}
	\label{tab_comp_srh}
	\centering
	\begin{tabular}{lrrrrr}
		\hline
		Model            & \multicolumn{1}{c}{Dice$\uparrow$}  & \multicolumn{1}{c}{PR$\uparrow$} & \multicolumn{1}{c}{RE$\uparrow$} & \multicolumn{1}{c}{NSD$\uparrow$} & \multicolumn{1}{c}{HD95$\downarrow$}   \\ \hline
		nnFormer         & 0.4291                   & 0.4574                        & 0.4856                     & 0.4455                  & 148.27                   \\
		UNETR++          & 0.4514                   & 0.6951                        & 0.4566                     & 0.4812                  & 308.67                   \\
		CoTr             & 0.5748                   & 0.6772                        & 0.5853                     & 0.6151                  & 117.78                   \\
		SwinUNETR        & 0.5970                   & 0.5751                        & 0.6725                     & 0.6184                  & 91.93                    \\
		SwinUNETR w/ SSL & 0.6052                   & 0.6219                        & 0.6830                     & 0.6371                  & 111.09                   \\
		nnUNet           & 0.6001                   & 0.6898                        & 0.6264                     & 0.6338                  & 162.00                   \\
		Ours             & \textbf{0.6935}          & \textbf{0.7270}               & \textbf{0.7275}            & \textbf{0.7428}         & \textbf{78.33}           \\ \hline
	\end{tabular}
\end{table}

\subsection{Ablation study and analysis}
In this section, we present and analyze the results of extensive ablation studies on the RJH dataset for each proposed component. During the ablation process, we progressively enhance the framework by 1) data modality augmentation, 2) AG-sampling, 3) AG-SSL and 4) AF-loss, corresponding to the following four subsections, respectively. 

\subsubsection{Efficacy of data modality augmentation}
To study the effect of introducing the auto-generated organ masks as extra data modalities, we perform experiments that ablate different input channels. As listed in Table \ref{tab_modality}, 
when combining CT with MOS masks, over 2\% improvement in Dice is observed compared with the CT-only baseline; when further incorporating the OOI modality, the model achieves the best scores in most metrics such as Dice, precision, recall and NSD, which surpasses the CT baseline by a large margin (e.g., $\sim$4.9\% in Dice and $\sim$5.5\% in NSD), demonstrating the effectiveness of employing these two anatomy-aware data modalities in model input. 

\begin{table}[]
	\caption{Model performance with different input modalities.}
	\label{tab_modality}
	\centering
	\begin{tabular}{lrrrrr}
		\hline
		Input modalities & \multicolumn{1}{c}{Dice$\uparrow$}  & \multicolumn{1}{c}{PR$\uparrow$} & \multicolumn{1}{c}{RE$\uparrow$} & \multicolumn{1}{c}{NSD$\uparrow$} & \multicolumn{1}{c}{HD95$\downarrow$}   \\ \hline
		CT               & 0.6001                   & 0.6898                        & 0.6264                     & 0.6338                  & 162.00                   \\
		CT + MOS         & 0.6243                   & 0.6778                        & 0.6603                     & 0.6630                  & \textbf{119.81}          \\
		CT + MOS + OOI   & \textbf{0.6490}          & \textbf{0.7062}               & \textbf{0.6864}            & \textbf{0.6899}         & 126.34                   \\ \hline
	\end{tabular}
\end{table}

\subsubsection{Efficacy of AG-sampling}
To investigate whether the proposed anatomy-guided sampling strategy boosts segmentation performance, we compare it with a range of heuristic alternatives, such as uniform sampling, sampling from certain organs, hybrid sampling combining foreground (i.e., the tumor region, FG) and background or certain organs, etc. We implement 7 different sampling strategies (including ours) in total, and the results are listed in Table \ref{tab_smpl}. All these experiments are conducted with CT+OOI+MOS input modalities. 
\begin{table}[]
	\caption{Model performance with different sampling strategies.}
	\label{tab_smpl}
	\centering
	\begin{tabular}{lrrrrr}
		\hline
		Sampling strategy                       & \multicolumn{1}{c}{Dice$\uparrow$}  & \multicolumn{1}{c}{PR$\uparrow$} & \multicolumn{1}{c}{RE$\uparrow$} & \multicolumn{1}{c}{NSD$\uparrow$} & \multicolumn{1}{c}{HD95$\downarrow$}   \\ \hline
		Uniform                                 & 0.6320                   & 0.7011                        & 0.6636                     & 0.6730                  & \textbf{112.50}          \\
		Uniform colon               & 0.6114                   & \textbf{0.7996}               & 0.6035                     & 0.6674                  & 138.92                   \\
		Uniform OOI                             & 0.6433                   & 0.7696                        & 0.6633                     & 0.6820                  & 169.08                   \\
		Hybrid uniform & 0.6490                   & 0.7062                        & \textbf{0.6864}            & 0.6899                  & 126.34                   \\
		Hybrid colon                    & 0.5945                   & 0.7441                        & 0.6015                     & 0.6350                  & 174.53                   \\
		Hybrid OOI                      & 0.6473                   & 0.7415                        & 0.6706                     & 0.6945                  & 154.40                   \\
		AG sampling (ours)           & \textbf{0.6616}          & 0.7569                        & 0.6734                     & \textbf{0.7103}         & 115.40                   \\ \hline
	\end{tabular}
\end{table}

From Table \ref{tab_smpl}, we first observe that the basic uniform sampling strategy can achieve reasonable results (0.6320 in Dice), the uniform OOI strategy (sampling uniformly from OOI) leads to better performance (0.6433 in Dice), while the colon sampling scheme (sampling patches uniformly from colon) produces worse results (0.6114 in Dice). Then, inspired by the nnUNet framework where additional 33\% patches are sampled from the foreground region (i.e., the tumor, FG), we further try three hybrid sampling schemes that combine FG oversampling with the aforementioned schemes, denoted as ``Hybrid uniform'' (67\% patches are sampled uniformly and 33\% from FG, which the same as the nnUNet setting), ``Hybrid colon'' (67\% patches from colon and 33\% from FG) and ``Hrbrid OOI'' (67\% patches from OOI and 33\% from FG) in Table \ref{tab_smpl}, respectively. We can observe that these several heuristic methods can hardly achieve better results than the nnUNet's default sampling setting. On the other hand, by examining the proposed anatomy-guided sampling, we observe a significant performance improvement in Dice (0.6616 vs 0.6490) and NSD (0.7103 vs 0.6899) metrics over the nnUNet baseline, and it also ranks high on other metrics. These observations suggest that, first, simply sampling from certain organ voxels may not be a better strategy than the conventional foreground oversampling scheme; second, the proposed anatomy-guided sampling scheme can surpass all the compared alternatives (including the strong nnUNet baseline) since it can produce an optimal sampling map (in terms of the gain function in (\ref{eq_sampling})) by taking into consideration multiple relevant factors such as proximity to high-value regions (i.e., the OOI region), sample diversity, patch size, and foreground emphasis. 

\subsubsection{Efficacy of anatomy-guided self-supervised learning}
To examine the effectiveness of the proposed AG-SSL task, we compare it with two standard mask-based SSL tasks, i.e., image inpainting \cite{tang2022self} (also denoted as image inner-cutout in \cite{zhou2021models}) and masked auto-encoder (MAE)\footnote{Though MAE is initially proposed for vision transformers, here we implement it with UNet.} \cite{he2022masked}. For the inpainting strategy, up to 50\% voxels in random-sized blocks can be dropped, following a similar protocol to the masked volume inpainting in \cite{tang2022self}; for MAE, dropping patch size is set as $4\times 32\times 32$ (about $20\times 25\times 25$ mm), and 50\% of them are randomly dropped; for AG-SSL, we always drop all voxels in the bowel wall region as described in \S\ref{sec_ssl}. For all these methods, the dropped voxels are replaced with random Gaussian noise. Besides, we only apply the masks to the CT channel without changing the OOI and MOS channels. 
\begin{table}[]
	\caption{Model performance with different self-supervision tasks.}
	\label{tab_ssl}
	\centering
	\begin{tabular}{lrrrrr}
		\hline
		Pretraining           & \multicolumn{1}{c}{Dice$\uparrow$}  & \multicolumn{1}{c}{PR$\uparrow$} & \multicolumn{1}{c}{RE$\uparrow$} & \multicolumn{1}{c}{NSD$\uparrow$} & \multicolumn{1}{c}{HD95$\downarrow$}   \\ \hline
		From scratch          & 0.6616                   & \textbf{0.7569}               & 0.6734                     & 0.7103                  & 115.40                   \\
		Image inpaint         & 0.6729                   & 0.7200                        & 0.7036                     & 0.7166                  & 100.48                   \\
		MAE& 0.6605                   & 0.7372                        & 0.6836                     & 0.7095                  & 101.72                   \\
		AG-smpl (ours) & \textbf{0.6900}          & 0.7163                        & \textbf{0.7308}            & \textbf{0.7354}         & \textbf{85.16}           \\ \hline
	\end{tabular}
\end{table}

As listed in Table \ref{tab_ssl}, we first observe that the inpainting and MAE schemes are both beneficial to the model performance. For example, the inpainting scheme improves the segmentation Dice by about 1\%, and both methods improve the surface distance metrics HD95 by some margin. By employing the proposed AG-SSL strategy, we further observe considerable performance improvement on most measurements, where nearly 3\% improvement in Dice is achieved over the unpreptrained baseline, and it also achieves the best scores on all the evaluation metrics except precision. These results clearly evidence the effectiveness of the anatomy-inspired SSL strategy, which can help the model understand the high-risk bowel wall regions during the pretraining stage.

\subsubsection{Efficacy of anatomy-focalized loss}
To further study how the AF-loss function influences model performance, we conduct ablation studies on AF-loss under pretrained and unpretrained scenarios\footnote{Note that we do NOT employ the AF-loss scheme during pretraining.}, as shown in Table \ref{tab_mask}. 
\begin{table}[]
	\caption{Model performance whether employing anatomy-focalizing loss, under different pretraining statuses.}
	\label{tab_mask}
	\centering
	\begin{tabular}{cclllll}
		\hline
		Pretrain   & AF-loss  & \multicolumn{1}{c}{Dice$\uparrow$}  & \multicolumn{1}{c}{PR$\uparrow$} & \multicolumn{1}{c}{RE$\uparrow$} & \multicolumn{1}{c}{NSD$\uparrow$} & \multicolumn{1}{c}{HD95$\downarrow$}   \\ \hline
		\ding{55}  & \ding{55}  & 0.6616                   & 0.7569                        & 0.6734                     & 0.7103                  & 115.40                   \\
		\ding{55}  & \checkmark & \textbf{0.6963}          & \textbf{0.7413}               & 0.7238                     & 0.7384                  & 91.14                    \\
		\checkmark & \ding{55}  & 0.6900                   & 0.7163                        & \textbf{0.7308}                     & 0.7354                  & 85.16                    \\
		\checkmark & \checkmark & 0.6935                   & 0.7270                        & 0.7275            & \textbf{0.7428}         & \textbf{78.33}           \\ \hline
	\end{tabular}
\end{table}
First, when the model is trained from scratch, applying the AF-loss function brings substantial performance improvement (the first two rows in Table \ref{tab_mask}), where the overall segmentation Dice is increased from 0.6616 to 0.6963, and all other metrics are also consistently improved by a considerable margin,
clearly evidencing the effectiveness of the AF-loss strategy. 
Second, when the model is initialized by pretrained weights, the AF-loss scheme is still helpful (the last two rows in Table \ref{tab_mask}), though the margin is relatively small, within 1\% on metrics like Dice and NSD. The different effects under the two circumstances can be explained by the fact that both the AF-loss and the AG-SSL strategies try to help the model understand and focus on the correct region (i.e., the general gastro-intestinal area, especially the ``organ wall'' tissues near the borders), thus once the model has been pretrained with the AG-SSL strategy, it already has the capacity of paying sufficient attention to the bowel wall area, thus the effect of further employing masked loss function becomes marginal. Overall, we observe that the model equipped with AF-loss yields the better results in the ablation studies, verifying its effectiveness. 

Moreover, we summarize the ablation process in Table \ref{tab_summary} encompassing the baseline implementation and the final rows of Table \ref{tab_modality}$\sim$\ref{tab_mask}, which clearly reflects the progressively refined CRC segmentation performance, where M, P, S, F denote data \underline{m}odality augmentation, AG-sam\underline{p}ling, AG-\underline{S}SL and A\underline{F}-loss, respectively.

\begin{table}[]
	\caption{Ablation summary.}
	\label{tab_summary}
	\centering
	\begin{tabular}{ccccrrrrr}
		\hline
		\makebox[5pt]M& \makebox[5pt]P&\makebox[5pt]S& \makebox[5pt]F& \multicolumn{1}{c}{Dice$\uparrow$}  & \multicolumn{1}{c}{PR$\uparrow$} & \multicolumn{1}{c}{RE$\uparrow$} & \multicolumn{1}{c}{NSD$\uparrow$} & \multicolumn{1}{c}{HD95$\downarrow$}   \\ \hline
		& && & 0.6001                   & 0.6898                        & 0.6264                     & 0.6338                  & 162.00                   \\
		\checkmark& && & 0.6490& 0.7062& 0.6864& 0.6899& 126.34                   \\
		\checkmark       & \checkmark& & & 0.6616& \textbf{0.7569}                        & 0.6734                     & 0.7103&115.40                   \\
		\checkmark       & \checkmark&\checkmark& & 0.6900& 0.7163                        & \textbf{0.7308}            & 0.7354         & 85.16           \\
		\checkmark       & \checkmark&\checkmark& \checkmark& \textbf{0.6935}                   & 0.7270                        & 0.7275            & \textbf{0.7428}         & \textbf{78.33}           \\ \hline
	\end{tabular}
\end{table}

\subsection{Qualitative results}
\subsubsection{Colorectal cancer segmentation results}
In Fig. \ref{fig_quality_msd}, we present some representative CRC segmentation results of the proposed method and two strong competitors, i.e., nnUNet and SwinUNETR. It can be observed that the predictions of the proposed framework resemble the ground truth well in most cases, while the other two competitors tend to miss some tumor voxels. This observation is consistent with the quantitative results, where the proposed model usually has better Dice and recall scores. 

\newcommand\widx{0.16}
\newcommand\wid{0.13}
\newcommand\hs{0.01pt}
\newcommand\qlmsd{quality/msd}
\newcommand\qlrjh{quality/rjh}
\begin{figure*}
	\captionsetup[subfigure]{labelformat=empty}
	\centering
	
	\subfloat{\includegraphics[width=\wid\textwidth]{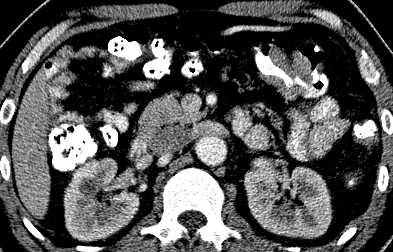}}\hspace{\hs}
	\subfloat{\includegraphics[width=\wid\textwidth]{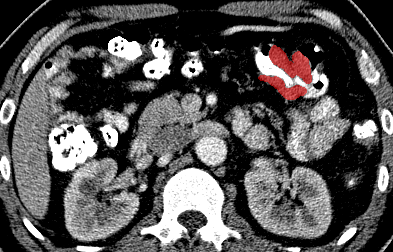}}\hspace{\hs}
	\subfloat{\includegraphics[width=\wid\textwidth]{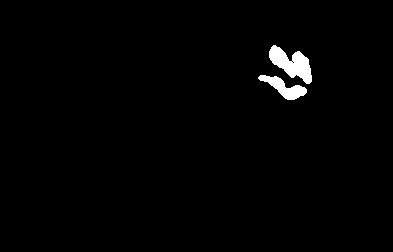}}\hspace{\hs}
	\subfloat{\includegraphics[width=\wid\textwidth]{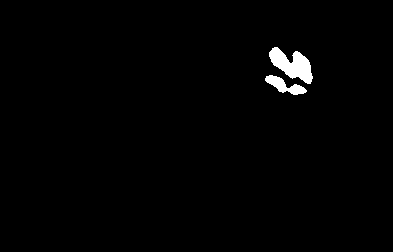}}\hspace{\hs}
	\subfloat{\includegraphics[width=\wid\textwidth]{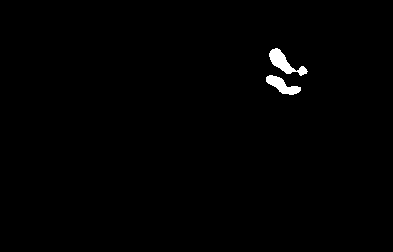}}\hspace{\hs}
	\subfloat{\includegraphics[width=\wid\textwidth]{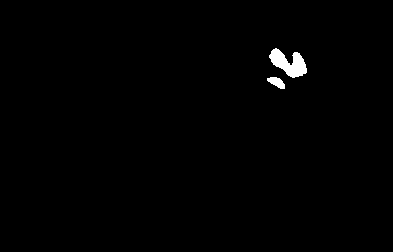}}
	\\[-0.1ex]
	\vspace{-4pt}
	\subfloat{\includegraphics[width=\wid\textwidth]{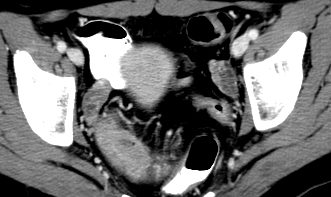}}\hspace{\hs}
	\subfloat{\includegraphics[width=\wid\textwidth]{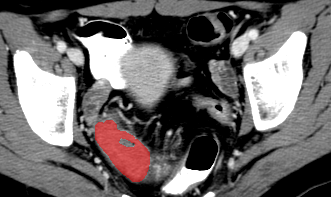}}\hspace{\hs}
	\subfloat{\includegraphics[width=\wid\textwidth]{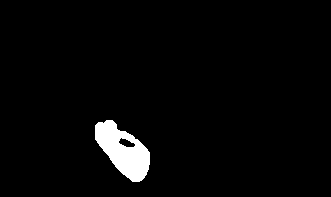}}\hspace{\hs}
	\subfloat{\includegraphics[width=\wid\textwidth]{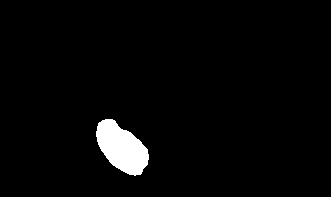}}\hspace{\hs}
	\subfloat{\includegraphics[width=\wid\textwidth]{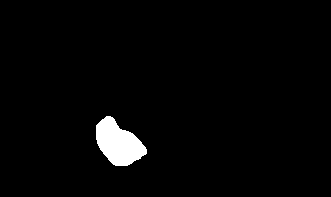}}\hspace{\hs}
	\subfloat{\includegraphics[width=\wid\textwidth]{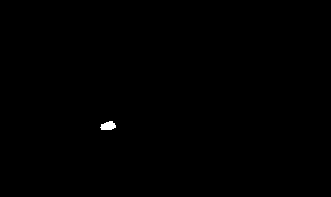}}
	\\[-0.1ex]
	\vspace{-4pt}
	\subfloat{\includegraphics[width=\wid\textwidth]{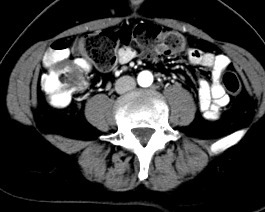}}\hspace{\hs}
	\subfloat{\includegraphics[width=\wid\textwidth]{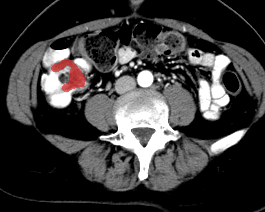}}\hspace{\hs}
	\subfloat{\includegraphics[width=\wid\textwidth]{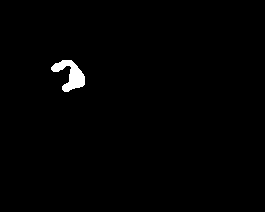}}\hspace{\hs}
	\subfloat{\includegraphics[width=\wid\textwidth]{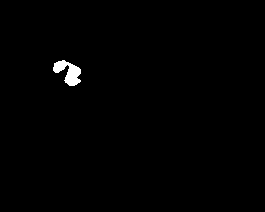}}\hspace{\hs}
	\subfloat{\includegraphics[width=\wid\textwidth]{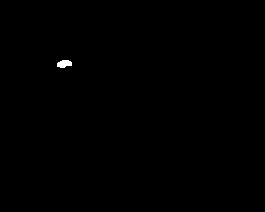}}\hspace{\hs}
	\subfloat{\includegraphics[width=\wid\textwidth]{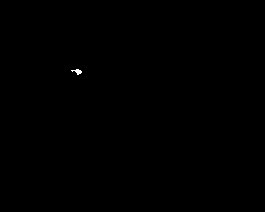}}
	\\[-0.1ex]
	\vspace{-4pt}
	\subfloat{\includegraphics[width=\wid\textwidth]{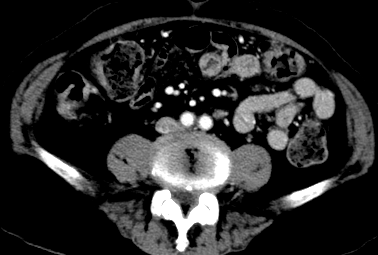}}\hspace{\hs}
	\subfloat{\includegraphics[width=\wid\textwidth]{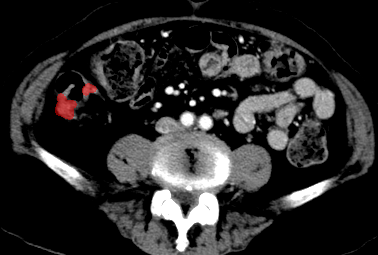}}\hspace{\hs}
	\subfloat{\includegraphics[width=\wid\textwidth]{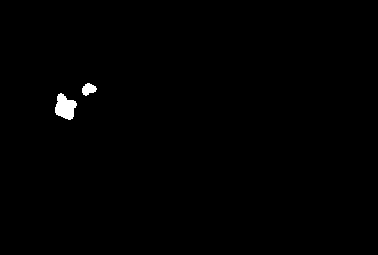}}\hspace{\hs}
	\subfloat{\includegraphics[width=\wid\textwidth]{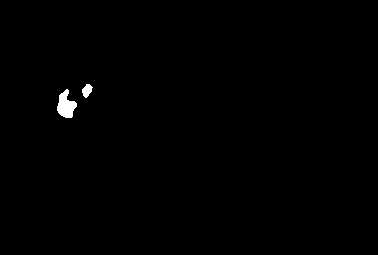}}\hspace{\hs}
	\subfloat{\includegraphics[width=\wid\textwidth]{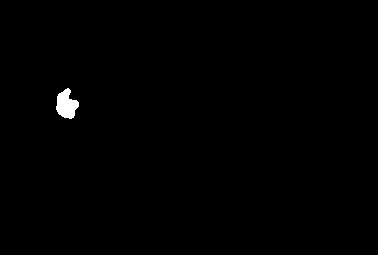}}\hspace{\hs}
	\subfloat{\includegraphics[width=\wid\textwidth]{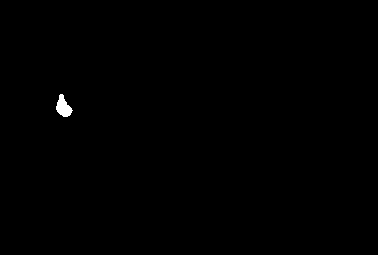}}
	\\[-0.1ex]
	\vspace{-4pt}
	\subfloat{\includegraphics[width=\wid\textwidth]{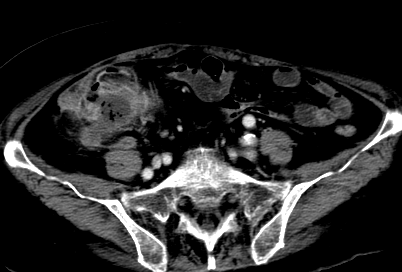}}\hspace{\hs}
	\subfloat{\includegraphics[width=\wid\textwidth]{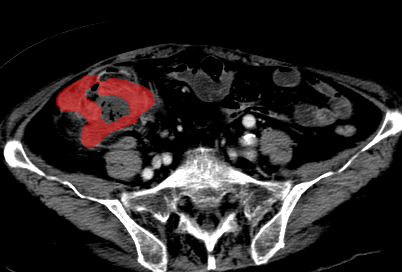}}\hspace{\hs}
	\subfloat{\includegraphics[width=\wid\textwidth]{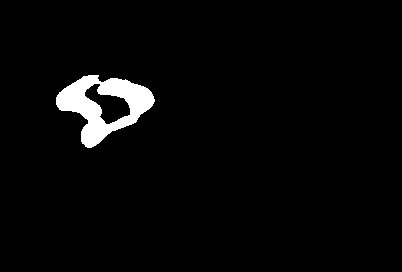}}\hspace{\hs}
	\subfloat{\includegraphics[width=\wid\textwidth]{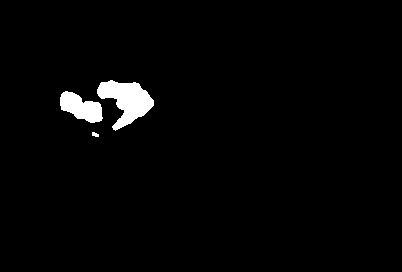}}\hspace{\hs}
	\subfloat{\includegraphics[width=\wid\textwidth]{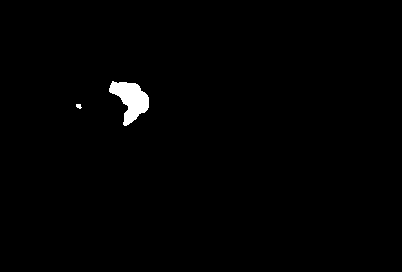}}\hspace{\hs}
	\subfloat{\includegraphics[width=\wid\textwidth]{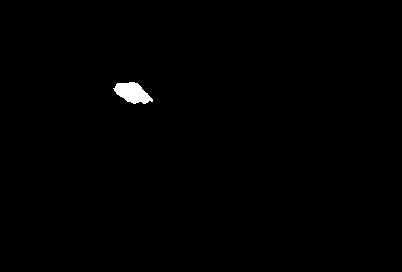}}
	\\[-0.1ex]
	\vspace{-4pt}
	
	\subfloat[CT]{\includegraphics[width=\wid\textwidth]{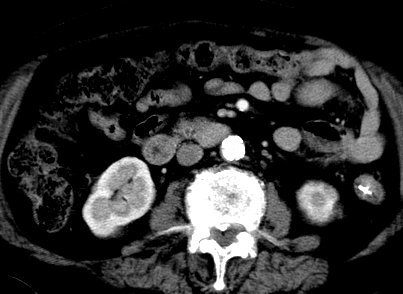}}\hspace{\hs}
	\subfloat[GT-overlap]{\includegraphics[width=\wid\textwidth]{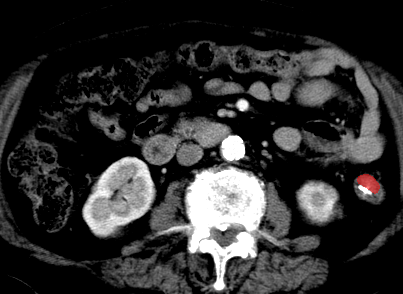}}\hspace{\hs}
	\subfloat[GT]{\includegraphics[width=\wid\textwidth]{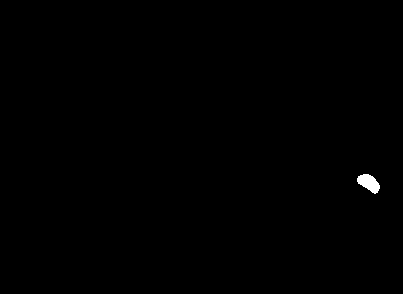}}\hspace{\hs}
	\subfloat[Ours]{\includegraphics[width=\wid\textwidth]{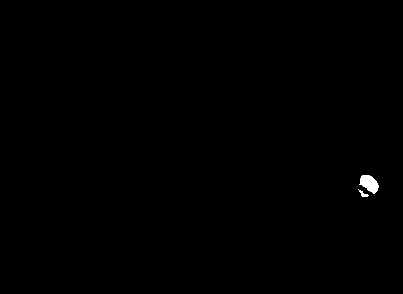}}\hspace{\hs}
	\subfloat[nnUNet]{\includegraphics[width=\wid\textwidth]{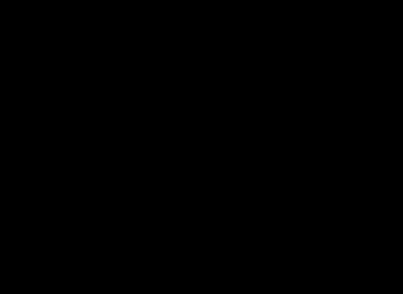}}\hspace{\hs}
	\subfloat[SwinUNETR]{\includegraphics[width=\wid\textwidth]{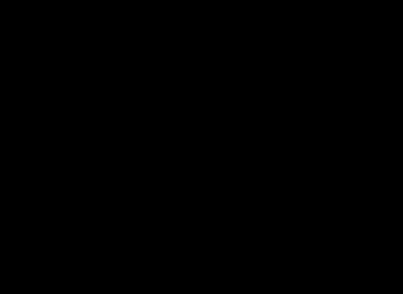}}
	
	\caption{Qualitative segmentation results of several exemplar testing cases (top three scans are from the MSD-Colon and bottom three ones are from the in-house RJH dataset). For clear visualization and correspondence, we first plot the CT scans and the overlapped GTs in the first two columns, followed by the binary GT masks and binary predicted masks of different methods.}
	\label{fig_quality_msd}
\end{figure*}

\subsubsection{Visualization of sampled locations}
To give the readers more intuitions on different sampling strategies, we record the locations that are sampled during training and visualize them in Fig. \ref{fig_smpl}. From this figure, we observe that the proposed AG-sampling strategy achieves a better balance among foreground, the OOI region, and voxel diversity when compared with the alternative methods. In particular, we notice that a key difference between AG-sampling and nnUNet's sampling is that nnUNet's samples concentrate on the foreground region densely, while our strategy tends to pick more diverse locations around the tumor region, which is made possible by the regularization term in the objective of (\ref{eq_sampling}) who encourages patch diversity.
%

\newcommand\widy{0.09}
\newcommand\dsmpl{smpl}
\begin{figure}
	\captionsetup[subfigure]{labelformat=empty}
	\centering
	
	\subfloat[OOI]{\includegraphics[width=\widy\textwidth]{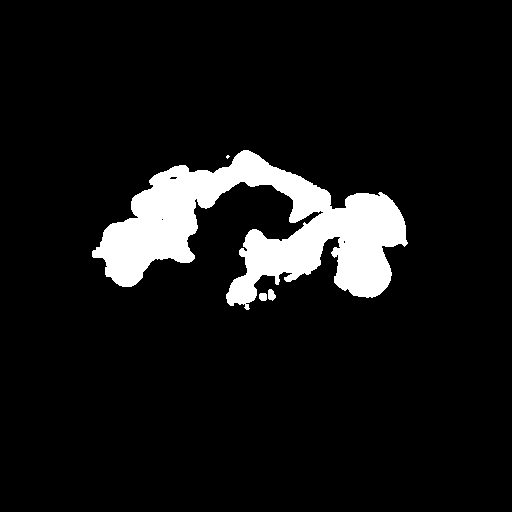}}\hspace{\hs}
	\subfloat[Tumor]{\includegraphics[width=\widy\textwidth]{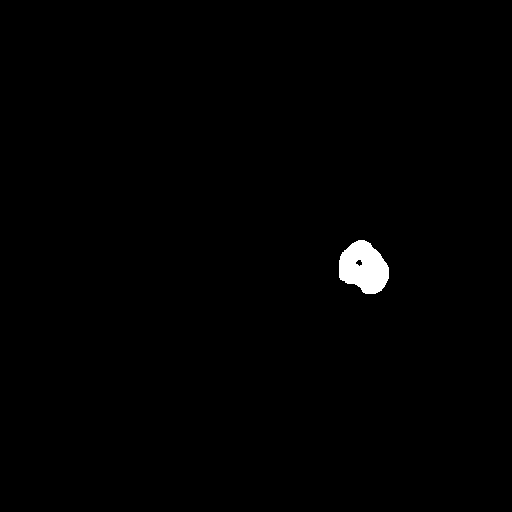}}\hspace{\hs}
	\subfloat[PSM]{\includegraphics[width=\widy\textwidth]{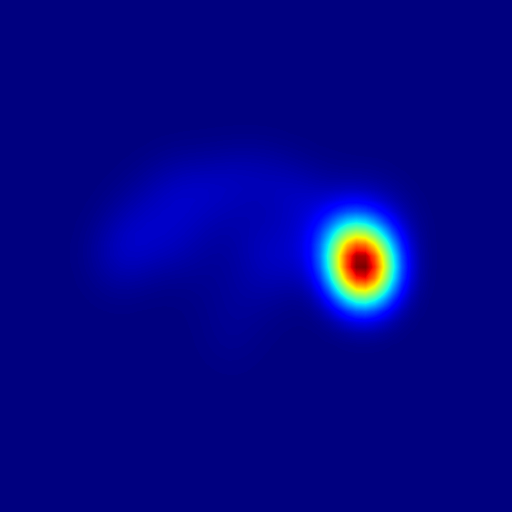}}\hspace{\hs}
	\subfloat[Uniform]{\includegraphics[width=\widy\textwidth]{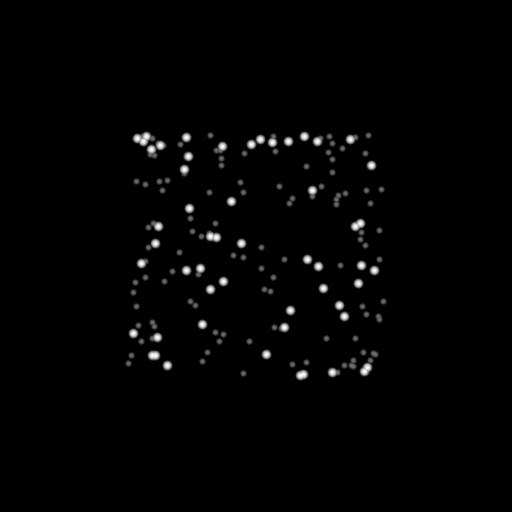}}\hspace{\hs}
	\subfloat[Uni. OOI]{\includegraphics[width=\widy\textwidth]{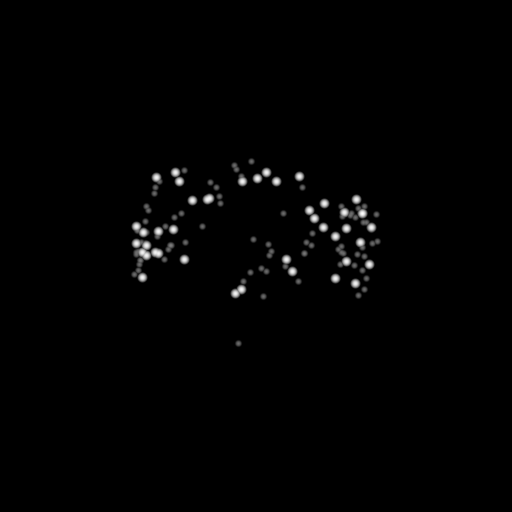}}\hspace{\hs}
	\\[-0.1ex]
		\vspace{-6pt}
	\subfloat[Uni. Colon]{\includegraphics[width=\widy\textwidth]{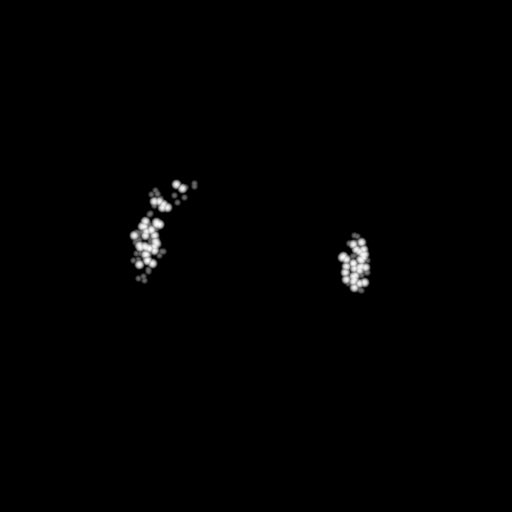}}\hspace{\hs}	
	\subfloat[Hyb. OOI]{\includegraphics[width=\widy\textwidth]{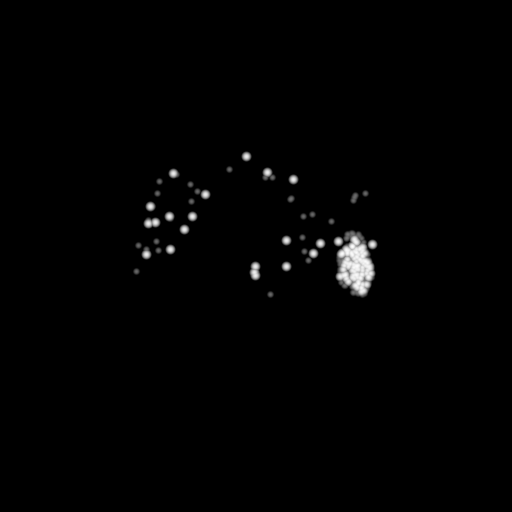}}\hspace{\hs}
	\subfloat[Hyb. Colon]{\includegraphics[width=\widy\textwidth]{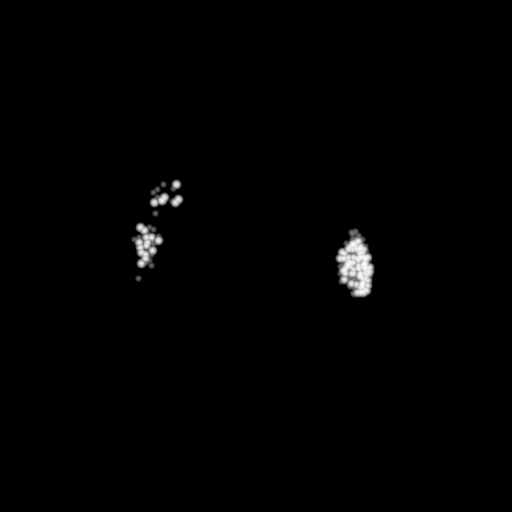}}\hspace{\hs}
	\subfloat[nnUNet]{\includegraphics[width=\widy\textwidth]{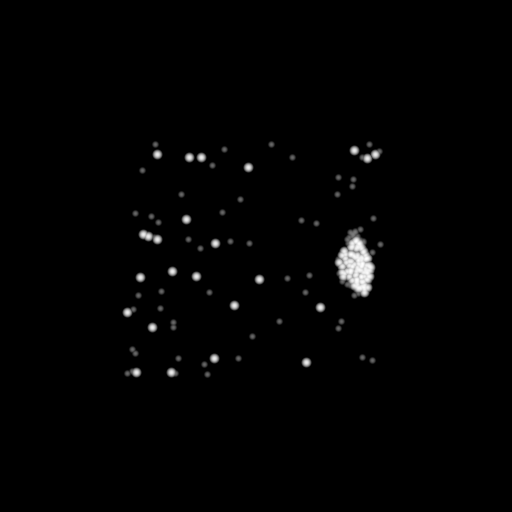}}\hspace{\hs}
	\subfloat[Ours]{\includegraphics[width=\widy\textwidth]{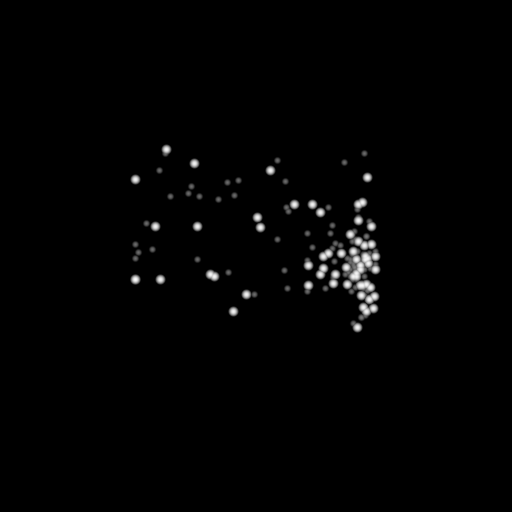}}\hspace{\hs}
	
	\caption{Point clouds of the sampled patches' centers for different sampling strategies. Each point is represented by a small spheroid (instead of a single voxel) for clear visualization.}
	\label{fig_smpl}
\end{figure}

\newcommand\discuss{discuss}

\section{Discussion and Conclusion}
In this paper, we have presented a novel anatomy-guided colorectal cancer segmentation framework that effectively leverages the auto-generated anatomical information to boost the segmentation performance. First, we utilize existing multi-organ segmentation models to obtain rich anatomical information from CT scans without expensive manual annotation. Then, we build a training pipeline that is seamlessly integrated with the auto-generated imperfect anatomical knowledge across its data feeding, training patch sampling, self-supervised pertaining, and supervised learning phases. Extensive experiments are conducted on two colorectal cancer segmentation datasets, where the proposed framework surpasses a range of 3D medical image segmentation algorithms by a large margin, and detailed ablation studies further verify the efficacy of each proposed component. 
As the proposed framework is inherently generic and can be smoothly adapted to other tasks, we hope that this work could inspire more studies that leverage anatomical information to guide medical image segmentation. 

Despite the aforementioned advantages, the proposed framework is still limited in several aspects. First, the framework only considers image features for CRC segmentation, while information from other modalities may provide further help. Second, the current model treats all types of CRC as a single category, though it is more desired to identify detailed cancer phenotypes in clinical practices.

In the future, to address the limitations in processing cross-modality information and in dealing with detailed cancer phenotypes, we plan to further improve this framework from two aspects. First, we will equip our model with multi-modality abilities by leveraging the power of large language models (LLM), so that it can utilize cross-modality information associated with the same patient, such as demographic data, genomic information, pathology report, diagnosis text, patients' chief complaint, etc. Second, we plan to collect wider-spectral cancer data such as cancer staging, cancer typing, prognosis information, etc., thus the model could deliver much more clinical values. 

\bibliography{bibfile}

\begin{thebibliography}{10}

\bibitem{sung2021global}
H.~Sung, J.~Ferlay, R.~L. Siegel, M.~Laversanne, I.~Soerjomataram, A.~Jemal,
  and F.~Bray, ``Global cancer statistics 2020: Globocan estimates of incidence
  and mortality worldwide for 36 cancers in 185 countries,'' {\em CA: a cancer
  journal for clinicians}, vol.~71, no.~3, pp.~209--249, 2021.

\bibitem{argiles2020localised}
G.~Argil{\'e}s, J.~Tabernero, R.~Labianca, D.~Hochhauser, R.~Salazar,
  T.~Iveson, P.~Laurent-Puig, P.~Quirke, T.~Yoshino, J.~Taieb, {\em et~al.},
  ``Localised colon cancer: Esmo clinical practice guidelines for diagnosis,
  treatment and follow-up,'' {\em Annals of Oncology}, vol.~31, no.~10,
  pp.~1291--1305, 2020.

\bibitem{ronneberger2015u}
O.~Ronneberger, P.~Fischer, and T.~Brox, ``U-net: Convolutional networks for
  biomedical image segmentation,'' in {\em Medical Image Computing and
  Computer-Assisted Intervention}, Springer, 2015.

\bibitem{isensee2021nnu}
F.~Isensee, P.~F. Jaeger, S.~A. Kohl, J.~Petersen, and K.~H. Maier-Hein,
  ``nnu-net: a self-configuring method for deep learning-based biomedical image
  segmentation,'' {\em Nature methods}, vol.~18, no.~2, pp.~203--211, 2021.

\bibitem{huang2018hl}
Y.-J. Huang, Q.~Dou, Z.-X. Wang, L.-Z. Liu, L.-S. Wang, H.~Chen, P.-A. Heng,
  and R.-H. Xu, ``Hl-fcn: Hybrid loss guided fcn for colorectal cancer
  segmentation,'' in {\em 2018 IEEE 15th international symposium on biomedical
  imaging (ISBI 2018)}, pp.~195--198, IEEE, 2018.

\bibitem{panic2020convolutional}
J.~Panic, A.~Defeudis, S.~Mazzetti, S.~Rosati, G.~Giannetto, L.~Vassallo,
  D.~Regge, G.~Balestra, and V.~Giannini, ``A convolutional neural network
  based system for colorectal cancer segmentation on mri images,'' in {\em 2020
  42nd Annual International Conference of the IEEE Engineering in Medicine \&
  Biology Society (EMBC)}, pp.~1675--1678, IEEE, 2020.

\bibitem{ye2021anatomy}
X.~Ye, D.~Guo, J.~Ge, X.~Di, Z.~Lu, J.~Xiao, G.~Yao, L.~Lu, D.~Jin, and S.~Yan,
  ``Anatomy guided thoracic lymph node station delineation in ct using deep
  learning model,'' {\em International Journal of Radiation Oncology, Biology,
  Physics}, vol.~111, no.~3, pp.~e120--e121, 2021.

\bibitem{yao2022deepcrc}
L.~Yao, Y.~Xia, H.~Zhang, J.~Yao, D.~Jin, B.~Qiu, Y.~Zhang, S.~Li, Y.~Liang,
  X.-S. Hua, {\em et~al.}, ``Deepcrc: Colorectum and colorectal cancer
  segmentation in ct scans via deep colorectal coordinate transform,'' in {\em
  Medical Image Computing and Computer Assisted Intervention}, Springer, 2022.

\bibitem{landman2015miccai}
B.~Landman, Z.~Xu, J.~Igelsias, M.~Styner, T.~Langerak, and A.~Klein, ``Miccai
  multi-atlas labeling beyond the cranial vault--workshop and challenge,'' in
  {\em Proc. MICCAI Multi-Atlas Labeling Beyond Cranial Vault—Workshop
  Challenge}, vol.~5, p.~12, 2015.

\bibitem{wasserthal2022totalsegmentator}
J.~Wasserthal, M.~Meyer, H.-C. Breit, J.~Cyriac, S.~Yang, and M.~Segeroth,
  ``Totalsegmentator: robust segmentation of 104 anatomical structures in ct
  images,'' {\em arXiv preprint arXiv:2208.05868}, 2022.

\bibitem{luo2022word}
X.~Luo, W.~Liao, J.~Xiao, J.~Chen, T.~Song, X.~Zhang, K.~Li, D.~N. Metaxas,
  G.~Wang, and S.~Zhang, ``Word: A large scale dataset, benchmark and clinical
  applicable study for abdominal organ segmentation from ct image,'' {\em
  Medical Image Analysis}, vol.~82, p.~102642, 2022.

\bibitem{shen2017deep}
D.~Shen, G.~Wu, and H.-I. Suk, ``Deep learning in medical image analysis,''
  {\em Annual review of biomedical engineering}, vol.~19, pp.~221--248, 2017.

\bibitem{cciccek20163d}
{\"O}.~{\c{C}}i{\c{c}}ek, A.~Abdulkadir, S.~S. Lienkamp, T.~Brox, and
  O.~Ronneberger, ``3d u-net: learning dense volumetric segmentation from
  sparse annotation,'' in {\em Medical Image Computing and Computer-Assisted
  Intervention}, Springer, 2016.

\bibitem{vaswani2017attention}
A.~Vaswani, N.~Shazeer, N.~Parmar, J.~Uszkoreit, L.~Jones, A.~N. Gomez,
  {\L}.~Kaiser, and I.~Polosukhin, ``Attention is all you need,'' {\em Advances
  in neural information processing systems}, vol.~30, 2017.

\bibitem{dosovitskiy2020image}
A.~Dosovitskiy, L.~Beyer, A.~Kolesnikov, D.~Weissenborn, X.~Zhai,
  T.~Unterthiner, M.~Dehghani, M.~Minderer, G.~Heigold, S.~Gelly, {\em et~al.},
  ``An image is worth 16x16 words: Transformers for image recognition at
  scale,'' in {\em International Conference on Learning Representations}, 2020.

\bibitem{liu2021swin}
Z.~Liu, Y.~Lin, Y.~Cao, H.~Hu, Y.~Wei, Z.~Zhang, S.~Lin, and B.~Guo, ``Swin
  transformer: Hierarchical vision transformer using shifted windows,'' in {\em
  Proceedings of the IEEE/CVF international conference on computer vision},
  pp.~10012--10022, 2021.

\bibitem{xie2021cotr}
Y.~Xie, J.~Zhang, C.~Shen, and Y.~Xia, ``Cotr: Efficiently bridging cnn and
  transformer for 3d medical image segmentation,'' in {\em Medical Image
  Computing and Computer Assisted Intervention}, pp.~171--180, Springer, 2021.

\bibitem{tang2022self}
Y.~Tang, D.~Yang, W.~Li, H.~R. Roth, B.~Landman, D.~Xu, V.~Nath, and
  A.~Hatamizadeh, ``Self-supervised pre-training of swin transformers for 3d
  medical image analysis,'' in {\em Proceedings of the IEEE/CVF Conference on
  Computer Vision and Pattern Recognition}, pp.~20730--20740, 2022.

\bibitem{zhou2023nnformer}
H.-Y. Zhou, J.~Guo, Y.~Zhang, X.~Han, L.~Yu, L.~Wang, and Y.~Yu, ``nnformer:
  Volumetric medical image segmentation via a 3d transformer,'' {\em IEEE
  Transactions on Image Processing}, 2023.

\bibitem{wang2018deep}
J.~Wang, J.~Lu, G.~Qin, L.~Shen, Y.~Sun, H.~Ying, Z.~Zhang, and W.~Hu, ``A deep
  learning-based autosegmentation of rectal tumors in mr images,'' {\em Medical
  physics}, vol.~45, no.~6, pp.~2560--2564, 2018.

\bibitem{jiang2021ala}
Y.~Jiang, S.~Xu, H.~Fan, J.~Qian, W.~Luo, S.~Zhen, Y.~Tao, J.~Sun, and H.~Lin,
  ``Ala-net: Adaptive lesion-aware attention network for 3d colorectal tumor
  segmentation,'' {\em IEEE transactions on medical imaging}, vol.~40, no.~12,
  pp.~3627--3640, 2021.

\bibitem{liu2019accurate}
X.~Liu, S.~Guo, H.~Zhang, K.~He, S.~Mu, Y.~Guo, and X.~Li, ``Accurate
  colorectal tumor segmentation for ct scans based on the label assignment
  generative adversarial network,'' {\em Medical physics}, vol.~46, no.~8,
  pp.~3532--3542, 2019.

\bibitem{lorenz2000generation}
C.~Lorenz and N.~Krahnst{\"o}ver, ``Generation of point-based 3d statistical
  shape models for anatomical objects,'' {\em Computer vision and image
  understanding}, vol.~77, no.~2, pp.~175--191, 2000.

\bibitem{heimann2009statistical}
T.~Heimann and H.-P. Meinzer, ``Statistical shape models for 3d medical image
  segmentation: a review,'' {\em Medical image analysis}, vol.~13, no.~4,
  pp.~543--563, 2009.

\bibitem{cuadra2004atlas}
M.~B. Cuadra, C.~Pollo, A.~Bardera, O.~Cuisenaire, J.-G. Villemure, and J.-P.
  Thiran, ``Atlas-based segmentation of pathological mr brain images using a
  model of lesion growth,'' {\em IEEE transactions on medical imaging},
  vol.~23, no.~10, pp.~1301--1314, 2004.

\bibitem{iglesias2015multi}
J.~E. Iglesias and M.~R. Sabuncu, ``Multi-atlas segmentation of biomedical
  images: a survey,'' {\em Medical image analysis}, vol.~24, no.~1,
  pp.~205--219, 2015.

\bibitem{hoogi2016adaptive}
A.~Hoogi, A.~Subramaniam, R.~Veerapaneni, and D.~L. Rubin, ``Adaptive
  estimation of active contour parameters using convolutional neural networks
  and texture analysis,'' {\em IEEE transactions on medical imaging}, vol.~36,
  no.~3, pp.~781--791, 2016.

\bibitem{hatamizadeh2019deep}
A.~Hatamizadeh, A.~Hoogi, D.~Sengupta, W.~Lu, B.~Wilcox, D.~Rubin, and
  D.~Terzopoulos, ``Deep active lesion segmentation,'' in {\em Machine Learning
  in Medical Imaging: 10th International Workshop, MLMI 2019, in Conjunction
  with MICCAI 2019}, Springer, 2019.

\bibitem{li2021agmb}
Y.~Li, G.~Zeng, Y.~Zhang, J.~Wang, Q.~Jin, L.~Sun, Q.~Zhang, Q.~Lian, G.~Qian,
  N.~Xia, {\em et~al.}, ``Agmb-transformer: Anatomy-guided multi-branch
  transformer network for automated evaluation of root canal therapy,'' {\em
  IEEE Journal of Biomedical and Health Informatics}, vol.~26, no.~4,
  pp.~1684--1695, 2021.

\bibitem{zhou2021anatomy}
B.~Zhou, Z.~Augenfeld, J.~Chapiro, S.~K. Zhou, C.~Liu, and J.~S. Duncan,
  ``Anatomy-guided multimodal registration by learning segmentation without
  ground truth: Application to intraprocedural cbct/mr liver segmentation and
  registration,'' {\em Medical image analysis}, vol.~71, p.~102041, 2021.

\bibitem{jiang2022anatomy}
M.~Jiang, Y.~Chen, J.~Yan, Z.~Xiao, W.~Mao, B.~Zhao, S.~Yang, Z.~Zhao,
  T.~Zhang, L.~Guo, {\em et~al.}, ``Anatomy-guided spatio-temporal graph
  convolutional networks (ag-stgcns) for modeling functional connectivity
  between gyri and sulci across multiple task domains,'' {\em IEEE Transactions
  on Neural Networks and Learning Systems}, 2022.

\bibitem{amin2017eighth}
M.~B. Amin, F.~L. Greene, S.~B. Edge, C.~C. Compton, J.~E. Gershenwald, R.~K.
  Brookland, L.~Meyer, D.~M. Gress, D.~R. Byrd, and D.~P. Winchester, ``The
  eighth edition ajcc cancer staging manual: continuing to build a bridge from
  a population-based to a more “personalized” approach to cancer staging,''
  {\em CA: a cancer journal for clinicians}, vol.~67, no.~2, pp.~93--99, 2017.

\bibitem{antonelli2022medical}
M.~Antonelli, A.~Reinke, S.~Bakas, K.~Farahani, A.~Kopp-Schneider, B.~A.
  Landman, G.~Litjens, B.~Menze, O.~Ronneberger, R.~M. Summers, {\em et~al.},
  ``The medical segmentation decathlon,'' {\em Nature communications}, vol.~13,
  no.~1, p.~4128, 2022.

\bibitem{Roth_Summers_2015}
H.~Roth, L.~Lu, A.~Seff, K.~M. Cherry, J.~Hoffman, S.~Wang, J.~Liu, E.~Turkbey,
  and R.~M. Summers, ``A new 2.5 d representation for lymph node detection in
  ct (ct lymph nodes),'' 2015.

\bibitem{clark2013cancer}
K.~Clark, B.~Vendt, K.~Smith, J.~Freymann, J.~Kirby, P.~Koppel, S.~Moore,
  S.~Phillips, D.~Maffitt, M.~Pringle, {\em et~al.}, ``The cancer imaging
  archive (tcia): maintaining and operating a public information repository,''
  {\em Journal of digital imaging}, vol.~26, pp.~1045--1057, 2013.

\bibitem{bilic2023liver}
P.~Bilic, P.~Christ, H.~B. Li, E.~Vorontsov, A.~Ben-Cohen, G.~Kaissis,
  A.~Szeskin, C.~Jacobs, G.~E.~H. Mamani, G.~Chartrand, {\em et~al.}, ``The
  liver tumor segmentation benchmark (lits),'' {\em Medical Image Analysis},
  vol.~84, p.~102680, 2023.

\bibitem{shaker2022unetr++}
A.~Shaker, M.~Maaz, H.~Rasheed, S.~Khan, M.-H. Yang, and F.~S. Khan, ``Unetr++:
  delving into efficient and accurate 3d medical image segmentation,'' {\em
  arXiv preprint arXiv:2212.04497}, 2022.

\bibitem{zhou2021models}
Z.~Zhou, V.~Sodha, J.~Pang, M.~B. Gotway, and J.~Liang, ``Models genesis,''
  {\em Medical image analysis}, vol.~67, p.~101840, 2021.

\bibitem{he2022masked}
K.~He, X.~Chen, S.~Xie, Y.~Li, P.~Doll{\'a}r, and R.~Girshick, ``Masked
  autoencoders are scalable vision learners,'' in {\em Proceedings of the
  IEEE/CVF conference on computer vision and pattern recognition},
  pp.~16000--16009, 2022.

\end{thebibliography}
\bibliographystyle{ieeetr}

\end{document}